\documentclass[%
 reprint,
superscriptaddress,
showpacs,preprintnumbers,nofootinbib,
 amsmath,amssymb,
 aps,
 prd,
 longbibliography,
]{revtex4-1}

\usepackage[utf8]{inputenc}
\usepackage{cancel}
\usepackage{accents} 
\usepackage{mciteplus,slashed}
\usepackage{amssymb,cancel,amsmath}
\usepackage{dcolumn}% Align table columns on decimal point
\usepackage{bm}% bold math 
\usepackage[caption=false]{subfig}
\usepackage{appendix}
\usepackage{physics}
\usepackage{enumerate}
\usepackage[T1]{fontenc}	
\usepackage{csvsimple}
\usepackage{hyperref}
\usepackage[section]{placeins}
\usepackage[capitalise]{cleveref}
\usepackage{booktabs,bookmark}
\usepackage{graphicx}
\usepackage{mathrsfs}
\graphicspath{{Figures/}}
\usepackage[utf8]{inputenc}
\usepackage[dvipsnames]{xcolor}
\usepackage[normalem]{ulem}

\newcommand{\be}{\begin{equation}}
\newcommand{\ee}{\end{equation}}
\newcommand{\ba}{\begin{align*}}
\newcommand{\ea}{\end{align*}}
\newcommand{\bpm}{\begin{pmatrix}}
\newcommand{\epm}{\end{pmatrix}}
\newcommand{\bea}{\begin{eqnarray}}
\newcommand{\eea}{\end{eqnarray}}

\newcommand{\benum}{\begin{enumerate}}
\newcommand{\eenum}{\end{enumerate}}
\newcommand{\bi}{\begin{itemize}}
\newcommand{\ei}{\end{itemize}}

\newcommand{\gsim}{\lower.7ex\hbox{$\;\stackrel{\textstyle>}{\sim}\;$}}
\newcommand{\lsim}{\lower.7ex\hbox{$\;\stackrel{\textstyle<}{\sim}\;$}}

\def\e{\mathrm{e}}

\begin{document}

%\title{Searching for millicharged particles and strongly interacting dark matter \\ with a cosmic ray beam dump and neutrino telescopes}

\title{New Constraints on Millicharged Particles from Cosmic-ray Production}

\author{Ryan Plestid}
\email{rpl225@uky.edu}
\affiliation{Department of Physics and Astronomy, University of Kentucky  Lexington, KY 40506, USA}
\affiliation{Theoretical Physics Department, Fermilab, Batavia, IL 60510,USA}
\affiliation{Perimeter Institute for Theoretical Physics, 31 Caroline St. N., Waterloo, Ontario N2L 2Y5, Canada}
\affiliation{\mbox{Department of Physics \& Astronomy, McMaster University, 1280 Main St. W., Hamilton, Ontario L8S 4M1, Canada}}

\author{Volodymyr Takhistov}
\email{vtakhist@physics.ucla.edu}
\affiliation{Department of Physics and Astronomy, University of California, Los Angeles \\ Los Angeles, California, 90095-1547, USA}

\author{Yu-Dai Tsai}
\email{ytsai@fnal.gov}
\affiliation{Theoretical Physics Department, Fermilab, Batavia, IL 60510,USA}
\affiliation{Cosmic Physics Center, Fermi National Accelerator Laboratory, Batavia, IL 60510, USA}

\author{\\Torsten Bringmann}
\email{torsten.bringmann@fys.uio.no}
\affiliation{Department of Physics, University of Oslo, Box 1048, N-0371 Oslo, Norway}

\author{Alexander Kusenko}
\email{kusenko@ucla.edu}
\affiliation{Department of Physics and Astronomy, University of California, Los Angeles \\ Los Angeles, California, 90095-1547, USA}
\affiliation{Kavli Institute for the Physics and Mathematics of the Universe (WPI), UTIAS\\
The University of Tokyo, Kashiwa, Chiba 277-8583, Japan}

\author{Maxim Pospelov}
\email{mpospelov@perimeterinstitute.ca}
\affiliation{School of Physics and Astronomy, University of Minnesota, Minneapolis, MN 55455, USA}
\affiliation{William I. Fine Theoretical Physics Institute, School of Physics and Astronomy, University of Minnesota, Minneapolis, MN 55455, USA}

\date{\today}

\begin{abstract}
%Production should not be pluralized in this context.
%\yt{Cool. Just checking.}

We study the production of exotic millicharged particles (MCPs) from cosmic ray-atmosphere collisions, which constitutes a permanent MCP production source for all terrestrial experiments. Our calculation of the MCP flux can be used to reinterpret existing limits from experiments such as MACRO and Majorana on an ambient flux of ionizing particles. Large-scale underground neutrino detectors are particularly favorable targets for the resulting MCPs. Using available data from the Super-K experiment, we set new limits on MCPs, which are the  best in sensitivity reach for the mass range $0.1 \lesssim m_{\chi} \lesssim 0.5$\,GeV, and which are competitive with accelerator-based searches for masses up to 1.5\,GeV. Applying these constraints to models where a sub-dominant component of dark matter (DM) is  fractionally charged allows us to probe parts of the parameter space that are challenging for  conventional  direct-detection DM experiments, independently of any assumptions about the DM abundance.  These results can be further improved with the next generation of large-scale neutrino detectors. 
\end{abstract}

%\pacs{13.40.Hq,60.Le,13.85.Tp,14.80.−j}% PACS, the Physics and 
% Classification Scheme.
%\keywords{Suggested keywords}%Use showkeys class option if 
%display desired

\preprint{\hfill FERMILAB-PUB-20-044-A-T}
\preprint{\hfill INT-PUB-20-004}
\preprint{\hfill IPMU20-0015}

\maketitle 

%\yt{In the intro, we have 3 stables in one paragraph and two of them are \emph{stable}. We need to simplify that.}
%\tb{I re-phrased and shortened; please check.}
%\yt{1)still missing refs. Check question marks\\ 2)Torsten's concern}

\section{Introduction \label{sec:1-intro}}

The remarkable success of the Standard Model (SM), along with null results for new physics at the LHC, 
strongly suggests that if new physics exists below the TeV scale it can only be weakly coupled to SM 
degrees of freedom. While nearly decoupled from 
the SM,  such a \emph{dark sector} would likely leave its strongest imprint on SM degrees of freedom 
commensurate with its own dynamical energy scales~\cite{Alexander:2016aln}. It is interesting to note that the MeV\,-\,GeV regime 
both contains many SM particles (e.g.\ muons, mesons, and nucleons) 
and hosts a number of 
persistent anomalies, including the anomalous magnetic moment of the 
muon~\cite{Bennett:2006fi,Jegerlehner:2009ry,Miller:2012opa}. 
Furthermore, this energy range is interesting from a phenomenological point of view as it allows for many novel and complementary search strategies that can be used to probe the dark sector. 
For instance, new physics can often be efficiently probed by fixed target 
experiments~\cite{Bjorken:2009mm,Essig:2013lka,Batell:2009di,Lees:2014xha,Essig:2010xa} with 
high intensity electron~\cite{Prinz:1998ua,Prinz:2001qz} and proton 
beams~\cite{deNiverville:2011it, Kahn:2014sra, Pospelov:2017kep, Magill:2018jla,Magill:2018tbb,Arguelles:2018mtc,Kelly:2018brz,Arguelles:2019xgp,Tsai:2019mtm} 
where dark sector particles can be produced either directly, or through decays of copious amounts of mesons; 
collider experiments are also useful probes, especially for higher mass particles with larger 
couplings where they typically provide the leading 
constraints~\cite{Davidson:2000hf,Ball:2016zrp,Haas:2014dda}. 
A less explored opportunity of discovering dark-sector particles is to consider their production in 
cosmic-ray interactions, and subsequent detection in large detectors
(see~e.g.~Ref.~\cite{Kusenko:2004qc,Yin:2009yt,Hu:2016xas,Bringmann:2018cvk,Arguelles:2019ziu,Coloma:2019htx,Alvey:2019zaa}).

Historically, the discovery of SM particles in the MeV-GeV regime (e.g.~pions~\cite{Lattes:1947mw} and muons \cite{Neddermeyer:1937md}) has harnessed cosmic rays (a proton beam) bombarding the upper atmosphere (a fixed target). The advantage of cosmic rays over accelerator-based fixed target experiments is that the ``beam'' is always on, there are almost no angular losses (as long as attenuation in rock and atmosphere can be neglected) because the cosmic-ray flux is isotropic, and the detectors located ``downstream'' can have significant size (e.g.\ IceCube~\cite{Abbasi:2008aa}, Super-Kamiokande~\cite{Abe:2013gga}, Hyper-Kamiokande~\cite{Abe:2018uyc}, JUNO~\cite{Djurcic:2015vqa}, DUNE~\cite{Abi:2018dnh}). This suggests that cosmic rays, coupled with neutrino telescopes serving as downstream detectors,  are a  powerful tool with which to probe the dark sector.

%Historically, several SM particles in the MeV\,-\,GeV regime, e.g., pions and muons, were discovered through cosmic rays bombarding the  upper atmosphere.  In fact, there is a clear link between dark-sector searches relying on cosmic rays and accelerator-based fixed-target probes: cosmic rays constitute a beam source that is always ``on'',  and the downstream detectors one can utilize include large neutrino detectors like IceCube~\cite{Abbasi:2008aa}, Super-Kamiokande~\cite{Abe:2013gga}, Hyper-Kamiokande~\cite{Abe:2018uyc}, JUNO~\cite{Djurcic:2015vqa}, and DUNE~\cite{Abi:2018dnh}, which could be much larger than the usual fixed-target detectors. In addition, there are almost no angular losses (as long as attenuation in soil and atmosphere can be neglected)  because the cosmic-ray flux is isotropic. 

%~\cite{Neddermeyer:1937md})
%~\cite{Lattes:1947mw}
%
%\ak{
%Neutrino telescopes (e.g.\ IceCube~\cite{Abbasi:2008aa},  Super-Kamiokande~\cite{Abe:2013gga}, Hyper-Kamiokande~\cite{Abe:2018uyc}, JUNO~\cite{Djurcic:2015vqa}, DUNE~\cite{Abi:2018dnh}) can detect exotic particles produced in the atmosphere by cosmic rays, providing a testing ground for dark sector models independent of cosmological assumptions.
%}

In this work, we calculate the flux of millicharged particles\footnote{%
Also known as charged massive particles (CHAMPs)  \cite{Chuzhoy:2008zy,Dunsky:2018mqs}.
} 
(MCPs, see e.g.~Ref.~\cite{Smith:1987me,Dobroliubov:1989mr,Golowich:1986tj,Babu:1993yh,Gninenko:2006fi}), 
$\chi$,  arising from meson decays in the upper atmosphere for $m_\chi$ in the few MeV to few GeV regime. 
For this, we adopt a {\it minimal} MCP model that is based on only two assumptions: 
\begin{enumerate}
    \item The new particle $\chi$ couples to the SM photon with a strength $Q_\chi=\epsilon\times e$; we remain agnostic as to the origin of this charge. 
    \item The new particle is \emph{stable}; this is a natural consequence if $Q_\chi$ is the smallest (non-zero) charge in the dark sector. 
\end{enumerate}
As these features 
are relatively generic, MCPs can be thought of as a useful representative example of a stable 
dark sector particle with which to benchmark the impact of neutrino telescopes. 
In particular, since we consider only primary production in what follows, our constraints apply (possibly conservatively) to \emph{any} model that satisfies the above two assumptions.

%MCPs represent a minimal model of a \emph{stable} dark sector candidate that can be produced 
%via meson decays, originating from proton-nucleon collisions in the upper atmosphere. 

In addition to being a useful benchmark model, MCPs are  of interest  because of their potential impact on 21cm cosmology (potentially explaining the EDGES  anomaly~\cite{Bowman:2018yin,Munoz:2018pzp,Barkana:2018qrx,Liu:2019knx}), and their natural 
appearance in models of light dark matter (DM) interacting with the SM via a massless dark 
photon~\cite{Holdom:1985ag,ArkaniHamed:2008qn,Pospelov:2008zw,Agrawal:2016quu}. 
Boosted millicharged DM can also potentially explain a reported excess in direct
detection experiments~\cite{Kurinsky:2020dpb,Robinson:2020lqx}. 
Running parallel to these more cosmological 
motivations, the lack of constraints in the few MeV\,-\,few GeV regime has also motivated the proposal 
of dedicated detectors such as MilliQan \cite{Ball:2016zrp,Haas:2014dda} and  FerMINI \cite{Kelly:2018brz}.  
%\footnote{The limit of MCP's emerges as you take the dark photon mass to zero, $m_V\rightarrow 0$, however a light dark photon introduces an additional light degree of freedom and with it more stringent bounds from cosmology; these bounds do not apply in the case of a ``pure'' MCP that does not originate from kinetic mixing.})
%
\begin{figure}
  \includegraphics[width=\linewidth]{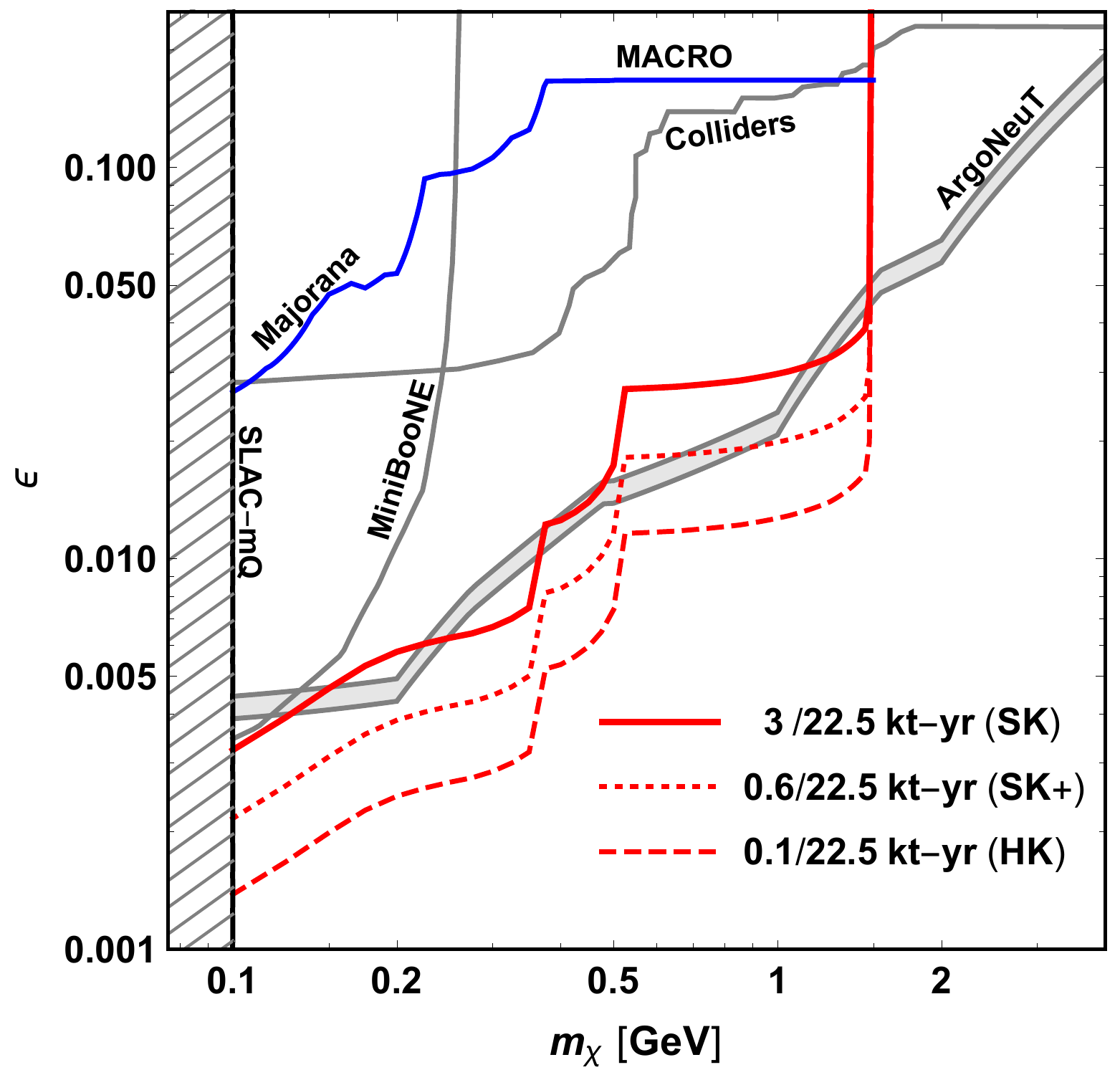}
  \caption{
  Exclusion limits for MCPs from cosmic-ray interactions (SK, red solid), obtained using analysis results 
  of the diffuse supernova neutrino background search in Super-K~\cite{Bays:2011si}, as well as 
  sensitivity projections for an improved SK analysis (SK+, red dotted) and near-future Hyper-K 
  (HK, red dashed). We also display new limits (blue) from recasting data of 
  MACRO~\cite{Ambrosio:2000kh,Ambrosio:2004ub} and Majorana~\cite{Alvis:2018yte}. Previous limits 
  from fixed target (SLAC MilliQ~\cite{Prinz:1998ua,Prinz:2001qz}, 
  MiniBooNE~\cite{Magill:2018tbb,Aguilar-Arevalo:2018wea}, ArgoNeuT~\cite{Acciarri:2019jly}) and 
  collider experiments~\cite{CMS:2012xi,Vogel:2013raa,Essig:2013lka,Jaeckel:2012yz} (as compiled 
  in Ref.~\cite{Acciarri:2019jly}) are shown for comparison. \label{money-plot} }
\end{figure}

One immediate consequence of our calculation of the cosmic ray-induced MCP flux is that existing 
bounds on a naturally occurring flux of MCPs can be converted into constraints on Lagrangian 
parameters $\epsilon$ and $m_\chi$, where the MCP charge is $Q_\chi=\epsilon \times e$. 
In fact, multiple such bounds already exist in the literature, but have never been translated into the 
$\epsilon-m_\chi$ plane because the relation between $\epsilon$, $m_\chi$ and the flux from cosmic 
rays had never been made explicit. Examples include constraints from 
MACRO~\cite{Ambrosio:2000kh,Ambrosio:2004ub}, Kamiokande-II~\cite{Mori:1990kw}, 
LSD~\cite{Aglietta:1994iv}, CDMS~\cite{Agnese:2014vxh} and Majorana~\cite{Alvis:2018yte}.  
Interestingly we find the resulting constraints to be roughly competitive with those from
existing  collider experiments, but sub-dominant to reported bounds from neutrino 
experiments~\cite{Magill:2018tbb,Aguilar-Arevalo:2018wea,Acciarri:2019jly}. 

Here we point out 
that neutrino telescopes can set new leading bounds on MCP couplings in the 
100\,MeV\,-\,500\,MeV regime based on existing data, surpassing the reach of fixed target experiments with 
neutrino detectors. We demonstrate this point explicitly by providing novel constraints 
based on published analyses by the Super-Kamiokande (Super-K, SK) collaboration 
searching for the  diffuse supernova neutrino background (DSNB)~\cite{Bays:2011si}. Our results, summarized in \cref{money-plot}, suggest that future neutrino telescopes could be able to act as 
the leading probe of MCPs in this mass regime.  
Furthermore, our results can be recast as a study of millicharged strongly interacting dark matter 
(SIDM)~\cite{1986PhLB..174..151G,Emken:2019tni}, allowing us to explore a region of interesting parameter 
space that cannot be easily studied by conventional underground direct-detection experiments (see \cref{SIDM} 
for a more detailed discussion).
%\tb{In PRL there are no sections, so the question is whether we also want to move Fig.~\ref{fig:SIDM} here
%to make it easier to refer to. In case we really go for PRL, the intro probably also needs some shortening.}
%\yt{Let's not worry about PRL at this stage, as discussed with Maxim and Volodymyr. }

Our study establishes neutrino telescopes as an important probe of the same MCP parameter 
space that motivated the proposal of MilliQan \cite{Ball:2016zrp,Haas:2014dda}  (and the similarly 
designed FerMINI~\cite{Kelly:2018brz}), studies on MCP bounds from neutrino 
experiments~\cite{Magill:2018tbb,Harnik:2019zee,Acciarri:2019jly}, and the proposed MCP DM 
explanation of the EDGES 21 cm anomaly~\cite{Munoz:2018pzp,Barkana:2018qrx,Liu:2019ogn}. 
In the context of MCP DM we emphasize that our constraints are independent of the fractional 
composition of composition of DM \cite{Dubovsky:2003yn,Dolgov:2013una,Kovetz:2018zan}, for other searches see e.g. Ref.\ \cite{Singh:2018von,Gninenko:2018ter,Liang:2019zkb,Liu:2019ogn}.
%\tb{Did you check this explicitly? I.e. is there no soil/atmosphere absorption even in the uppermost part
%of Fig.1?} \yt{I removed the attenuation statement because MACRO/MARJORANA curves may be affected indeed}
%
 Finally, the explicit calculation of the MCP flux from cosmic rays presented here will enable the use of neutrino telescopes as a robust platform for studying  MCPs, free from cosmological assumptions.  This has connections to charge quantization, which is itself connected to, but does not necessarily preclude \cite{Holdom:1985ag}, the existence of magnetic  monopoles~\cite{Dirac:1931kp}, Grand 
 Unification~\cite{Pati:1973uk,Georgi:1974sy,Georgi:1974my,Fritzsch:1974nn}, and quantum 
 gravity~\cite{Shiu:2013wxa}. 
 
 This paper is organized as follows. In Section \ref{sec:CR_meson}, we discuss the production of mesons from cosmic-ray collisions in the upper atmosphere.
 In Section \ref{sec:MCP_Meson}, the MCP flux from meson decays is calculated. We then discuss the detection of MCPs in neutrino telescopes in Section \ref{sec:MCP_lab}, and outline the kinematics of detecting MCPs. In Section \ref{sec:MCP_SIDM}, we discuss the millicharged SIDM and the constraints and projections that we can place based on our analysis.

%\ak{I am not sure what is meant by the sentence above. I don't think MCPs from kinetic mixing contradict the existence of monopoles. This should be explained or rephrased.}

%\yt{Certainly, that's why we say "connect to". It is possible to search for MCP in the case that dark photon is constrained by cosmology, means that the MCP is not from kinetic mixing.}

%Addressed 

%%%%%%%%%%%%%%%%%%%%%%%%%%%%%%%%%%%%%%%
%%%%%%%%%%%%%%%%%%%%%%%%%%%%%%%%%%%%%%%

\section{Cosmic-ray meson production \label{sec:CR_meson}}

Cosmic rays produce a sizeable number of mesons from interactions in the upper atmosphere, 
whose subsequent decay produces a continuous flux of MCPs. While the problem can be studied 
numerically with Monte Carlo simulations, we present here a semi-analytic treatment, allowing us to 
transparently illustrate the role of key ingredients.
Incoming cosmic rays are isotropically distributed on the sky, with the associated flux typically
quoted in terms of intensity, 
$[I_\text{CR}]=\text{GeV}^{-1}\text{cm}^{-2} \text{s}^{-1} \text{str}^{-1}$~\cite{Tanabashi:2018oca}. 
In our analysis we take this quantity as implemented in {\sf DarkSUSY}~\cite{Bringmann:2018lay}, 
based on Ref.~\cite{Boschini:2017fxq}, and focus on the dominant component of cosmic rays, free protons. 
For convenience, we will instead express the intensity in terms of the center-of-mass boost 
$\gamma_\text{cm}$ for CR protons impinging on atmospheric protons at rest and thus introduce 
$\mathcal{I}_\text{CR}(\gamma_\text{cm})= I_\text{CR}(E_p) \times \dd E_p/\dd \gamma_\text{cm}$,
where $\gamma_\text{cm}=\tfrac12\sqrt{s}/m_p$, $s$ is the Mandelstam variable for the $pp$ collision, and $m_p$ is the proton mass.

Taking into account that all incoming cosmic rays are eventually absorbed by the atmosphere, the 
amount of {\it primary} mesons $\mathfrak{m}$ produced in these collisions is approximately 
determined by the ratio of the 
inclusive cross section $\sigma_\mathfrak{m}$ for $pp\rightarrow \mathfrak{m} X$ with other particles 
$X$ to the total inelastic cross section for protons passing through atmospheric matter. 
We note that this is a rather conservative estimate for the {\it total} production of mesons given that
all final states in these primary interactions tend to trigger further cascades when interacting with the
atmosphere, resulting, among others, in a large multiplicity of (lower-energy) meson states. Here we neglect these contributions, which could be studied with a dedicated Monte Carlo simulations
of air showers.
%\tb{It's worth pointing out here that this a pretty conservative estimate because high-E CRs trigger 
%showers with enormous particle multiplicities: in this regime, the produced $X$ will act like a new 'primary'
%CR with  about half the initial energy. This implies that actual MC shower simulations
%are likely to improve our limits more than just marginally -- but also not by orders of magnitude, c.f.Fig.2.}
We model all interactions in the upper atmosphere as $pp$ collisions 
and therefore take the elastic cross section to be $\sigma_\text{in}(pp)$, whose dependence on 
$\gamma_\text{cm}$ is given in Ref.~\cite{Tanabashi:2018oca}. 
The resulting meson flux from 
cosmic-ray collisions in the upper atmosphere is then given by
\begin{equation}
  \Phi_\mathfrak{m}(\gamma_\mathfrak{m})= %\frac{2\pi}{4\pi} 
 \Omega_\text{eff}\int ~\mathcal{I}_\text{CR}(\gamma_\text{cm}) \frac{\sigma_{\mathfrak{m}}(\gamma_\text{cm})}{\sigma_\text{in}(\gamma_\text{cm})} P(\gamma_\mathfrak{m}|\gamma_\text{cm}) \,\dd  \gamma_\text{cm}\,,
\end{equation}
where $\Omega_\text{eff}\approx 2\pi$ is the effective solid angle from which MCPs can arrive
at the detector\footnote{By rescaling the muon's stopping power~\cite{Tanabashi:2018oca,Hu:2016xas}, we estimate that the energy loss of MCPs in the Earth's crust (standard rock) is roughly 50\,MeV/km for $\epsilon\sim 10^{-2}$.
While for the range of $\epsilon$ and energies that we are interested in here MCPs interact too strongly to penetrate the entire Earth, they are not significantly impeded to reach the detector when originating from the upper hemisphere.
}, 
and $P(\gamma_\mathfrak{m}|\gamma_\text{cm})$ represents the probability to get a meson
with boost $\gamma_\mathfrak{m}$ in the lab frame.
% from a $pp$ collision with a center of mass boost of $\gamma_\text{cm}$. 
%\tb{now introduced above}
The latter can be conveniently estimated (see Appendix~\ref{apps:mesboost}) 
from the differential production cross section with respect to %the longitudinal momentum fraction 
$x_F\equiv p_L/p_\text{max}$, %denoting the ``Feynman-$x$'' variable, 
where $p_L$ is the longitudinal momentum and $p_\text{max}$ is the maximum possible momentum:
\begin{equation}
  P(\gamma_\mathfrak{m}|\gamma_\text{cm})\approx 
 \sum_\alpha \frac{1}{\sigma_\mathfrak{m}} \times \dv{\sigma_\mathfrak{m}}{x_F}\times \dv{x^{(\alpha)}_F}{\gamma_\mathfrak{m}}\,.
\end{equation}
Here $\alpha=\pm$ denotes the two different possible contributions, see~\cref{feynman-x}, and 
$\dd \sigma_\mathfrak{m}/\dd x_F$ is a function of $\gamma_\text{cm}$ and $x_F(\gamma_\mathfrak{m})$. %Therefore, the required information to find the boost-distribution of a given meson species is $\sigma_\mathfrak{m}(\gamma_\text{cm})/\sigma_\text{in}(\gamma_\text{cm})$ and $\dd \sigma_\mathfrak{m} / \dd x_F$ as a function of $\gamma_\text{cm}$. 
%
%\begin{itemize}
%  \item The ratio between the production cross section, and the proton's inelastic cross section  as a function of $\gamma_\text{cm}$ (or equivalently $\sqrt{s}$)  $\sigma_\mathfrak{m}(\gamma_\text{cm})/\sigma_\text{in}(\gamma_\text{cm})$.
%  \item The shape of the differential cross section $\dd \sigma_\mathfrak{m} / \dd x_F$ as a function of $\gamma_\text{cm}$. 
%\end{itemize}
%

%Knowing the probability of a meson having $\gamma_\mathfrak{m}$ if it were produced, the next obvious quantity to consider is how many such mesons are acutally produced. If we consider only primary production of mesons (neglecting secondary and tertiary production from protons that are byproducts of previous inelastic collisions) then the number of mesons per-proton is given by  $\sigma_\mathfrak{m}/\sigma_\text{in}$ where $\sigma_\text{in}$ is the inclusive inelastic $pp$ cross section. 

The meson-production energy spectrum thus
depends on both the total meson cross section, $\sigma_\mathfrak{m}(\gamma_\text{cm})$, and the 
differential cross section with respect to $x_F$, or equivalently on 
$P(\gamma_\mathfrak{m}|\gamma_\text{cm})$. These quantities must be specified across 
a significant range of $\gamma_\text{cm}$ to reflect the large range of cosmic-ray energies, and we
do so by interpolating between existing data for selected values of fixed 
$\gamma_\text{cm}$  (see \cref{app:meson-prod}). Although both $\sigma_\mathfrak{m}$ 
and $P(\gamma_\mathfrak{m}|\gamma_\text{cm})$ influence the final resulting MCP flux, 
we find that the production cross section (which is also better measured) has a much stronger effect 
than the differential distribution.

\begin{figure}
    \includegraphics[width=\linewidth]{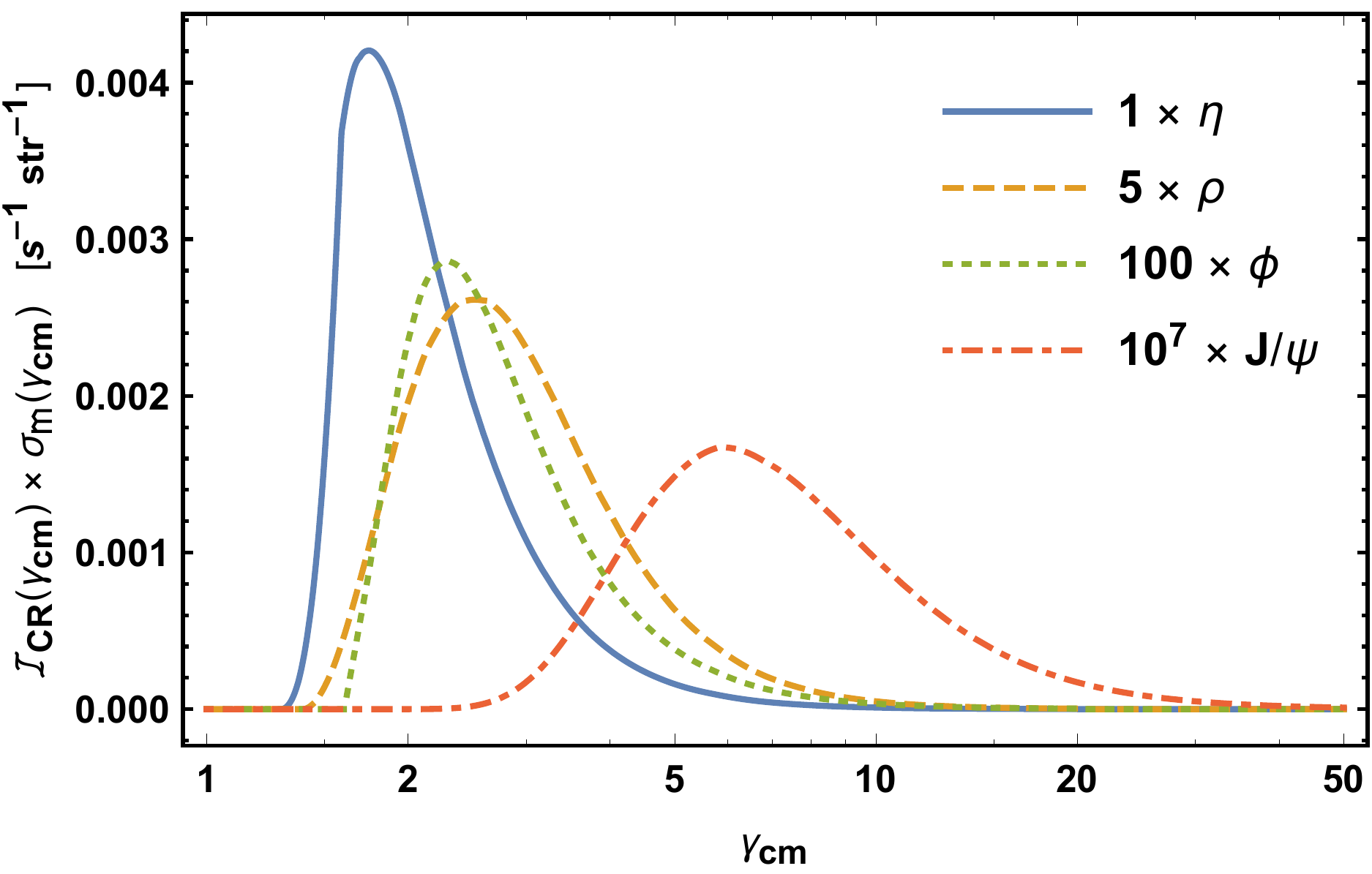}
    \caption{Differential cosmic-ray intensity multiplied by the meson production cross section as a function of $\gamma_\text{cm}$. For the resulting meson spectra see \cref{meson-fluxes} in \cref{app:MCPsignals}.
    %This gives us a rough handle on how many of each meson is produced, and what part of the cosmic ray spectrum dominantly contributes to that meson species' production.
    \label{production-modes} } 
\end{figure}

In principle, all possible mesons originating from $pp$ interactions and leading to MCPs (i.e.~those 
with substantial electromagnetic decay modes: $\pi$, $\eta$, $\omega$, $\rho$, $J/\psi$, $\Upsilon$ etc.) 
as well as direct production via Drell-Yan should be considered. For light MCPs produced via 
$\pi^0 \rightarrow \gamma \chi \bar{\chi}$ a combination of SLAC's milliQ experiment \cite{Prinz:1998ua} 
and LSND's search for electron-like scattering events \cite{Auerbach:2001wg} already strongly 
restricts the MCP parameter space~\cite{Magill:2018tbb}. 
%such that MCPs produced from cosmic rays are not expected to provide competitive sensitivity. 
We therefore restrict our discussion to the case of heavier MCPs, with  $m_\chi>\tfrac12 m_\pi$, 
where pion decay to MCPs is kinematically forbidden. 

To keep our discussion of meson production tractable, we focus on the dominant $\eta$, light vector, 
and $J/\psi$ mesons. While we have also quantitatively considered  $\Upsilon$ meson as well as direct 
Drell-Yan production, we found these contributions to be negligibly small (six orders of magnitude smaller than $J/\psi$) since in addition to smaller 
cross sections these processes require more energetic cosmic rays (with correspondingly much smallerfluxes). 

In \cref{app:meson-prod} we analyze and fit the available experimental data for $\eta$, $\rho,\omega,\phi$ 
and $J/\psi$, finding the total production cross section $\sigma_{pp\rightarrow \eta X}(\gamma_\text{cm})$ 
as well as the standard spectrum ``shape parameterization'' $\dd \sigma_\mathfrak{m} /\dd x_F$. 
In Fig.~\ref{production-modes} we display the resulting differential cosmic ray intensity 
$I_{\rm CR}(\gamma_{\rm cm})$ multiplied by $\sigma_\mathfrak{m}(\gamma_{\rm cm})$.
The shape of these curves is determined by the competition between a rising inclusive cross 
section and a sharply falling cosmic-ray flux, and illustrates which parts of the cosmic ray 
spectrum predominantly contributes to a given meson species.

%%%%%%%%%%%%%%%%%%%%%%%%%%%%%%%%%%%%%%%%%%%%%%%%%%%%
%%%%%%%%%%%%%%%%%%%%%%%%%%%%%%%%%%%%%%%%%%%%%%%%%%%%
%%%%%%%%%%%%%%%%%%%%%%%%%%%%%%%%%%%%%%%%%%%%%%%%%%%%
\section{MCP flux from meson decays \label{sec:MCP_Meson}
 }

Upon constructing $\Phi_\mathfrak{m}$ as outlined above, we can find the associated flux of MCPs 
from meson decays by folding the meson flux with the unit-normalized spectrum of MCPs in the lab frame, $P(\gamma_\chi|\gamma_\mathfrak{m})$,  and weighting by the decay branching ratio
\begin{equation}
  \Phi_\chi (\gamma_{\chi})= 2\sum_\mathfrak{m}  \text{BR}(\mathfrak{m}\rightarrow \chi\bar{\chi}) \int \dd \gamma_\mathfrak{m} \Phi_\mathfrak{m}(\gamma_\mathfrak{m}) P(\gamma_\chi|\gamma_\mathfrak{m})\,,
\end{equation}
where the factor of $2$ accounts for the contribution from both $\bar{\chi}$ and $\chi$. The quantity $P(\gamma_\chi|\gamma_\mathfrak{m})$ can be calculated from first principles, at leading order in $\epsilon$, as
\begin{equation}\label{p-gamma-chi}
  P(\gamma_\chi|\gamma_\mathfrak{m})= \left[\frac{1}{\Gamma} \dv{\Gamma}{\gamma_\chi}\right]_\text{lab}\,, 
\end{equation}
where $\Gamma$ is the decay rate for $\mathfrak{m}\rightarrow \chi\bar{\chi}$ and 
$\dd \Gamma/\dd \gamma_\chi$ is the differential rate with respect to the MCP boost, both evaluated 
in the lab frame (see Appendix~\ref{apps:mesdec}).

\begin{figure}
  \includegraphics[width=\linewidth]{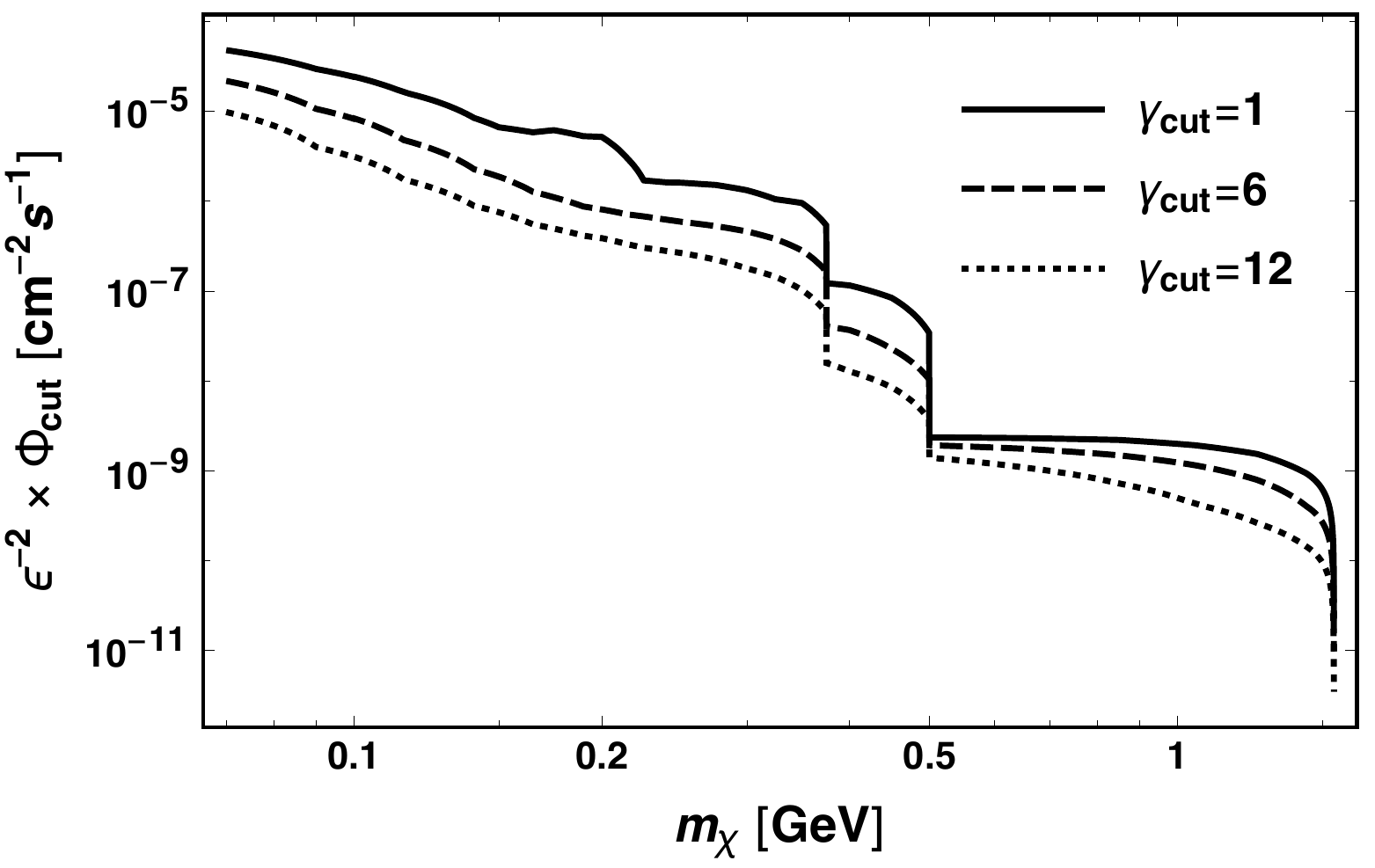}
  \caption{Fast-flux of MCPs $\Phi_\text{cut}$ due to meson decays as a function of MCP mass, 
  $m_\chi$, for three different choices of $\gamma_\text{cut}$. The spectrum for $\gamma_\text{cut}=1$ 
  is the full integrated MCP flux.  The meson mass thresholds are clearly visible, stemming from
 $\eta$, $\omega/\rho$, $\phi$, and finally $J/\psi$ (sequentially from left to right).
  \label{fastflux}} 
\end{figure}

Anticipating MCP detection, we define the integrated  ``fast-flux'' of MCPs satisfying $\gamma_\chi\geq \gamma_\text{cut}$ as
\begin{equation} \label{eq:fastflux}
  \Phi_\text{cut}(m_\chi, \gamma_\text{cut}) = \int_{\gamma_\text{cut}}^\infty  \dd \gamma_\chi ~\frac{\dd \Phi_\chi}{\dd \gamma_\chi}\,,
\end{equation}
where $\gamma_\text{cut}$ is set by the relevant experimental threshold.
In Fig.~\ref{fastflux} we display the mass-dependence of this quantity for several choices 
of $\gamma_\text{cut}$. The choice $\gamma_\text{cut}=1$ corresponds to the full integrated MCP flux, as relevant for low-threshold ionization experiments, while $\gamma_\text{cut}=6$ is adequate for experiments  with an electron recoil threshold of $T_\text{min}=16$\,MeV (as relevant for the physics analysis  of Super-K discussed below).
%

%%%%%%%%%%%%%%%%%%%%%%%%%%%%%%%%%%%%%%%%%%%%%%%%%%%%%%%%%%%%
\section{Detecting MCPs in laboratories \label{sec:MCP_lab}}

%Our computed MCP flux from cosmic-ray interactions establishes a permanent MCP production 
%source for all terrestrial experiments. 
As alluded to in the introduction, MCPs can deposit ionization energy directly within detectors,
which can be used as a probe of MCP couplings~\cite{Ambrosio:2000kh,Ambrosio:2004ub,Mori:1990kw,Aglietta:1994iv,Agnese:2014vxh,Alvis:2018yte}. 
Lacking an explicit calculation of cosmic rays as MCP source, previous searches have avoided  discussing the mass of the incident MCPs, and instead presented constraints on an ambient  MCP flux as a function of the fractional charge $\epsilon$. Our study allows us to directly translate these  results (and future searches) into limits on $\epsilon$ as a function of $m_\chi$, thus making direct  contact with Lagrangian parameters. We discuss the details of this translation  in \cref{apps:ionization}, and show our results in \cref{money-plot}.  We find that  ionization experiments are competitive with constraints from colliders around the 100 MeV  regime, but quickly become subdominant as $m_\chi$ is increased. Significant improvement  in detector exposure for ionization searches is expected in future experiments, and our results establish a quantitative baseline that can be used to estimate the potential future impact of upcoming projects such as LEGEND~\cite{Abgrall:2017syy}. 
%\vtx{; we speculate that ionization experiments may be able to surpass collider bounds in future generations of experiments.}   %Due to dramatically smaller detector exposure we estimate that these limits will be significantly less competitive for MCPs with $\epsilon \lesssim 10^{-1}$ than from large neutrino experiments that we discuss below, and hence we do not discuss them further\footnote{ Significant improvement in detector exposure for ionization searches is expected in future experiments such as LEGEND~\cite{Abgrall:2017syy}.}. We note that for large fractional charge (i.e. $\epsilon > 10^{-1}$) the effects of flux attenuation when propagating through medium, particularly through standard rock, become significant (see e.g.~\cite{Hu:2016xas}).  
We note that for MCPs with large charges of $\epsilon \gtrsim 10^{-1}$, as relevant for ionization searches, 
effects of attenuation when passing through Earth to reach typical detector depths of $\sim 1$\,km of 
standard rock (i.e.~few km water-equivalent) become significant (see e.g.~Ref.~\cite{Hu:2016xas}). 
Since we do not attempt a detailed translation of ionization bounds in this work, and this region is 
already well constrained by collider searches (which are not sensitive to attenuation), we do not 
consider these effects here.

Electron  scattering inside Cherenkov detectors, with recoils in the 10 MeV range, is a powerful probe  of MCPs~\cite{Magill:2018jla} (see also \cite{Hu:2016xas}). Counting electron-like events with recoil energies, $T_e'=2m_e(E_e-m_e)$, between $T_\text{min}$ and $T_\text{max}$ naturally introduces a windowed cross-section \cref{sigma-tilde} which can be well approximated (see \cref{cross-section-plot}) as
\begin{equation}\label{sigma-approx}
  \tilde{\sigma}_{e\chi}(\gamma_{\chi}) \approx  \frac{2\pi \alpha^2 \epsilon^2 }{2 T_\text{min} m_e } \qty( 1- \frac{T_\text{min}}{T_\text{max}})\Theta\qty(\gamma_\chi - \gamma_\text{cut})\,.
\end{equation}
Here, $\alpha$ is the fine-structure constant, $\Theta$ is the Heaviside step function and  $\gamma_\text{cut}\approx 0.6\sqrt{2 T_\text{min}/m_e}+0.4 \sqrt{2 T_\text{max}/m_e}$. The total resulting number of $\chi-e$ scattering events $N_{e\chi}$ for a given experiment is
\begin{align}\label{exact-signal}
  N_{e\chi} =&~ N_e \times t \times \int_{\gamma_\text{cut}}^\infty \dd \gamma_\chi  ~ \tilde{\sigma}_{e\chi}(\gamma_\chi) \frac{\dd \Phi_\chi}{\dd \gamma_\chi}(\gamma_\chi) \\
    \approx&~  N_e \times t\times   \frac{\pi \alpha^2 \epsilon^2 }{ T_\text{min} m_e } \qty( 1- \frac{T_\text{min}}{T_\text{max}}) \times \Phi_\text{cut}(m_\chi)\,, \notag
\end{align}
where $N_e$ is the number of electrons within the detector's fiducial volume and $t$ is the data collection period. % and we have used Eq.~\eqref{eq:fastflux}.

Using data sample and analysis results from the DSNB search in Super-K~\cite{Bays:2011si} we can 
place stringent new limits on MCPs from cosmic-ray production. This search looked at inverse beta 
decays $\bar{\nu}_e p \rightarrow n e^+$ with a positron recoil energy  %\footnote{Cherenkov detectors do not directly differentiate between electrons and positrons.} 
 $16\,\text{MeV}< T_{e^+}< 88\,\text{MeV}$, corresponding to $\gamma_\text{cut}\approx 6$,  effectively reducing the background from  cosmic ray muon spallation at lower energies (note that Cherenkov detectors do not directly differentiate between electrons and positrons). As detailed in \cref{apps:ionization}, the results of the likelihood analysis performed in  Ref.~\cite{Bays:2011si} (dashed curve  in their Fig.~19) can directly be employed to constrain the recoil electron spectrum from MCPs.

We show our resulting bounds on MCPs in \cref{money-plot}. The limits are competitive with  accelerator-driven searches across the $0.1 \lesssim m_{\chi} \lesssim 1.5$ GeV range. In the  range $0.1 \lesssim m_{\chi} \lesssim 0.5$ neutrino telescopes exceed the leading constraints from  both MiniBooNE and ArgoNeuT, demonstrating the potential of neutrino telescopes as a ``downstream'' detector. The results are quoted in terms of events per year per $22.5$\,kt of water, corresponding to the fiducial volume of Super-K as employed in Ref.~\cite{Bays:2011si}. 

Upcoming large neutrino experiments will be able to further improve on these results.  Additional background suppression due to improved neutron tagging will be possible in  an upcoming Super-K upgrade with gadolinium doping~\cite{Beacom:2003nk}, which we  denote as SK+ and assume a reach of $\sim0.6$ events/22.5 kt-yr in Fig.~\ref{money-plot}.  With a fiducial volume of 190\,kt the near future Hyper-K water Cherenkov  experiment~\cite{Abe:2018uyc} can further improve significantly on results of Super-K.  In Fig.~\ref{money-plot} we indicate this by assuming a year-long exposure and a sensitivity (in terms of SK's fiducial volume) of $\sim0.1$ events/22.5 kt-yr.

Other near future experiments with sizable fiducial volumes, such as DUNE  (40 kton, liquid argon)~\cite{Abi:2018dnh} and JUNO (20 kton, liquid scintillator)~\cite{Djurcic:2015vqa},  will complement water-based Cherenkov detectors as probes of the DSNB~\cite{Moller:2018kpn}, and hence can also serve as probes of atmospherically-produced MCPs. The solar and  spallation backgrounds at low energies are expected to be present in DUNE's DSNB  search~\cite{Zhu:2018rwc}, with an expected resulting energy cut-off of $\sim 20$ MeV for  a search as in SK. Due to favorable detector configuration and application of pulse-shape  discrimination techniques JUNO can perform DSNB search over a wider energy  range~\cite{An:2015jdp}, down to $\sim10$ MeV, albeit with overall statistics still considerably lower than that of Hyper-K.

%%%%%%%%%%%%%%%%%%%%%%%%%%%%%%%%%%%%%%%%%%%%%%%%%%%%%%%%%%%%
%%%%%%%%%%%%%%%%%%%%%%%%%%%%%%%%%%%%%%%%%%%%%%%%%%%%%%%%%%%%

%%%%%%%%%%%%%%%%%%%
\section{Strongly-Interacting DM \label{SIDM} \label{sec:MCP_SIDM} }

%\yt{UNDER CONSTRUCTION. Will be done Saturday morning.
%Need to add Neff ref. Unify the usage of ""}

\begin{figure}[t]
  \includegraphics[width=\linewidth]{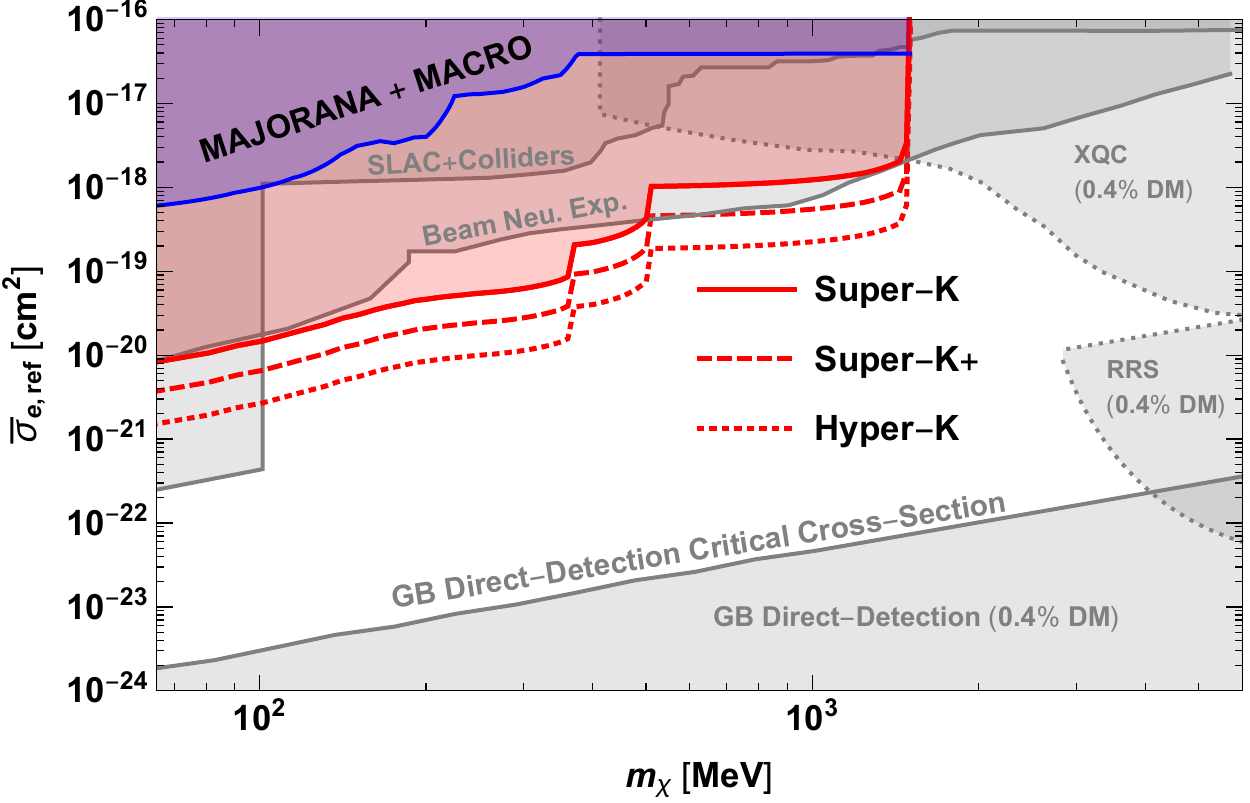}
  \caption{
  Constraints and sensitivity reaches that can cover the millicharged SIDM window, including our new bounds from recasting Super-K data \cite{Bays:2011si} (red) and the projection for Super-K (red, dashed) and Hyper-K (red, dotted). We also show the new bounds (blue) from recasting data of MACRO \cite{Ambrosio:2000kh,Ambrosio:2004ub} and Majorana \cite{Alvis:2018yte}.   Existing accelerator-based constraints \cite{Prinz:1998ua,Prinz:2001qz,CMS:2012xi,Jaeckel:2012yz,Essig:2013lka,Vogel:2013raa,Magill:2018tbb,Acciarri:2019jly} and direct-detection limits \cite{Rich:1987st,Erickcek:2007jv,Mahdawi:2018euy,Emken:2019tni} % (assuming 0.4$\%$ DM local abundance following Ref.~\cite{Emken:2019tni}) 
  are also shown. 
%\yt{Torsten also suggested: Moving Fig. 4 to the front?}
  \label{fig:SIDM}}
\end{figure}

A major focus of DM studies is the direct detection of DM particles using terrestrial detectors \cite{Goodman:1984dc}, typically placed underground. However, these searches  depend on the local flux of the DM particles of interest that could reach the experiment.  It has long been noted that when the DM-SM particle (mostly nuclei and electrons) cross section is large enough, this flux would be significantly attenuated~\cite{1986PhLB..174..151G,Starkman:1990nj}. The class of DM models featuring such large interactions with ordinary matter is often referred to as strongly interacting DM (SIDM\footnote{This is not identical with ``self-interacting DM'' \cite{Spergel:1999mh}.}).

In Ref.~\cite{Emken:2019tni},  DM-SM interactions through a dark photon kinetically mixing with $U(1)_Y$ are studied, focusing on the terrestrial effects on direct detection experiments.  In this case, DM scattering with electrons becomes more important than scattering with nuclei, so this is what we focus on in the following. Millicharged DM with cross sections larger than a critical value would have its average energy attenuated and be unable to trigger a detectable signature in ground-based direct-detection experiments~\cite{Emken:2019tni}. Above this critical cross section, there is a window of available parameter space where MCPs could constitute a sub-dominant component of DM ($\lesssim 0.4\%$ to avoid cosmological constraints~\cite{Dubovsky:2003yn,Dolgov:2013una,Kovetz:2018zan,Green:2019glg}), from hereon referred to as the millicharged SIDM window. New balloon and satellite experiments have been recently proposed \cite{Emken:2019tni} to  further explore this window, which could accommodate interesting DM models that could potentially explain the EDGES anomaly~\cite{Bowman:2018yin,Munoz:2018pzp,Berlin:2018sjs,Barkana:2018qrx,Liu:2019knx}.

In this section, we recast our bounds and projections on MCPs to explore this SIDM window. The results  in Fig.~\ref{fig:SIDM} are shown in terms of a ``reference cross section'' typically employed for direct detection experiments to compare the sensitivity reaches of different experiments. For millicharged DM, this is given by\footnote{In Ref.~\cite{Emken:2019tni}, the DM millicharge is generated a coupling to a massless dark photon that kinetically mixes with the SM  $U(1)_{\rm Y}$; here we directly consider the DM to have a millicharge under $U(1)_{\rm Y}$, with minimal theoretical assumptions about the origin of this charge. The reference cross section we consider corresponds to $\alpha_D=\alpha$ and $m_A' \rightarrow 0$ in Eq.~(2.6) of Ref.~\cite{Emken:2019tni}.}
\begin{equation}\label{eq:ref_cross}
\bar{\sigma}_{\rm e,ref}
= \frac{16 \pi \alpha^2 \epsilon^2 \mu^2_{\chi e}}{q^4_{ \rm d, ref}}\,,
\end{equation} 
where $\mu_{\chi e}$ is the reduced mass of the electron and $\chi$ and $q_{\rm d, ref}$ is the typical momentum transfer in $\chi-e$ scattering for semiconductor or noble-liquid targets from the local DM flux, taken to be $\alpha m_e$ \cite{Emken:2019tni}. Above the ``ground-based (GB) direct detection critical cross section'', one can see a regime enclosed by bounds from accelerator-based experiments~\cite{Prinz:1998ua,Prinz:2001qz,CMS:2012xi,Jaeckel:2012yz,Essig:2013lka,Vogel:2013raa,Magill:2018tbb,Acciarri:2019jly}, constraints from the above-atmosphere detector (RRS) \cite{Rich:1987st}, a rocket experiment 
(XQC)~\cite{Erickcek:2007jv,Mahdawi:2018euy}, and underground direct detection experiments~\cite{Emken:2019tni}.
% (assuming the millicharged SIDM to be 0.4$\%$ of the full DM 
%abundance to avoid stringent cosmological 
%constraints~\cite{Dubovsky:2003yn,Dolgov:2013una,Kovetz:2018zan}.)
%/\tb{this is already mentioned two paragraphs above}
We plot the bound of Super-K and sensitivity reaches for Super-K+, and Hyper-K. We do not consider bounds based on the MCP acceleration from astrophysical sources~\cite{Hu:2016xas,Dunsky:2018mqs,Li:2020wyl,Chuzhoy:2008zy}, since they rely on additional assumptions beyond local DM abundance. The constraint on the ultralight dark-photon mediator is also not shown since it is not an essential ingredient for minimal MCPs.

Our results in Fig. \ref{fig:SIDM} establish new constraints on the millicharged SIDM window. It is important to note that our bounds and projections are independent of any assumption about which fraction of the DM is millicharged.  Further, for reference cross sections below approximately $10^{-17}\,\mathrm{cm}^2$ our results are insensitive to attenuation in the Earth, given that cosmic-ray produced MCPs have much higher energy than that 
of the local DM flux.

%\yt{The attenuating of cosmic-ray produced MCP is only important when $\epsilon>0.1$, as discussed in Sec. \ref{}, which is already covered by accelerator-based searches.}

%\vt{We estimate that local DM particles of average speed $v \sim 10^{-3}$ experience significant energy losses passing through standard rock even for $\epsilon < 10^{-1}$, when attempting to reach typical detector depths of few kilometers of water-equivalent.}

%\yt{PRELIMINARY:The cosmic-ray produced MCP has averaged energy of ?? GeV to ?? GeV, while the local dark matter flux has energy of the level of ?? in this mass range. Thus, the energy loss of millicharged SIDM passing through earth's atmosphere and crust is much more significant for the local flux. Estimation shows that for the comic-ray produced MCP, only epsilon > 0.1 the attentuation becomes important. Since that regime is already covered by the existing accelerator-based bounds, we simply plot the MACRO+MAJO without considering attenuation (how abut making MAJO+MACRO dashed)}

\section{Summary}

We have considered MCP production from standard cosmic rays interacting with the atmosphere. This closes a gap in the MCP literature and constitutes a permanent MCP production source for all terrestrial experiments. We presented the first translation of long-standing bounds on an ambient MCP flux into bounds on the MCP charge $\epsilon$ as a function of its mass $m_\chi$, and demonstrated that large-scale underground neutrino experiments are particularly well suited for probing previously inaccessible parameter space. Using existing limits from Super-K's DSNB search we have placed new limits on MCPs for $0.1 \lesssim m_{\chi} \lesssim 1.5$ GeV, which  for $m_{\chi} \lesssim 0.5$ GeV exceed the sensitivity of fixed target experiments such as MiniBooNE and ArgoNeuT. These new limits are highly relevant also in scenarios where MCPs constitute an SIDM component because they are {\it i)} independent of the DM fraction made of such MCPs and {\it ii)} probe a part of the parameter space that cannot be readily tested with conventional direct-detection experiments. The results presented here will be further improved with upcoming large-scale neutrino experiments, and, since we only consider primary meson production, can likely be 
further strengthened by a more detailed modeling of cosmic-ray showers.

\section{Acknowledgements} 

We would like to thank Dr. T.-T. Yu for a stimulating discussion of the cosmic-ray-generated light dark matter flux. R.P. \& Y.-D.T. thank the University of Washington and the Institute for Nuclear theory for its hospitality during the final portion of this work. R.P.\ also thanks the Fermilab theory group for their hospitality and support. Y.-D.T. would like to thank KICP, University of Chicago, for the hospitality and support. A.K. and V.T. were  supported  by the U.S. Department of Energy (DOE) Grant No.  DE-SC0009937. A.K. was also supported by the World Premier International Research Center Initiative (WPI), MEXT Japan. R.P.\ was supported by an the Government of Canada through an NSERC PGS-D award, and by the U.S. Department of Energy, Office of Science, Office of High Energy Physics, under Award Number DE-SC0019095. This manuscript has been authored by Fermi Research Alliance, LLC under Contract No. DE-AC02-07CH11359 with the U.S. Department of Energy, Office of Science, Office of High Energy Physics. This work was partly performed at the Aspen Center for Physics, which is supported by National Science Foundation grant PHY-1607611. Research at the Perimeter Institute is supported in part by the Government of Canada through NSERC and by the Province of Ontario through MEDT.
 
%~\clearpage

\appendix 

\section{Cosmic-ray production kinematics}

\subsection{Boost of produced mesons}
\label{apps:mesboost}
The lab-frame energy of a meson produced in a collision, $E_\mathfrak{m}$, can be written as 
$E_\mathfrak{m} = \gamma_\text{cm} \mathcal{E} + \gamma_\text{cm} \beta_\text{cm} \mathcal{P}_\parallel$, 
where curly script variables refer to center of mass frame quantities. 
We can re-write this expression in terms of $x_F=\mathcal{P}_\parallel/p_\text{max}$ (``Feynman-x''), 
where $p_\text{max}= \frac12\sqrt{s}(1-m_\mathfrak{m}^2/s)$ is the largest possible longitudinal 
momentum allowed by kinematic constraints; $x_F$ therefore varies from $-1$ (backwards pointing) 
to $+1$ (forward pointing). Written in terms of $x_F$ our formula is given by  
$E_\mathfrak{m} = \gamma_\text{cm} p_\text{max} ( \mathcal{E}/p_\text{max}  + \beta_\text{cm} x_F)$, 
or %dividing through by $m_\mathfrak{m}$, 
\begin{equation}
  \gamma_\mathfrak{m} = \gamma_\text{cm} \frac{p_\text{max}}{m_\mathfrak{m}} \qty( \sqrt{x_F^2 + \frac{p_T^2}{p_\text{max}^2} + \frac{m_\mathfrak{m}^2}{p_\text{max}^2} } + \beta_\text{cm} x_F )\,.
\end{equation}
% 
%Since we are primarily interested in high-energetic MCPs resulting from meson decay, boosted above some energy threshold, our focus is on high meson boosts.  Hence, we can conservatively neglect the  transverse momentum contribution, which only serves to increase the speed of the mesonic progenitors. Furthermore, 
Since $p_T=\mathcal{P}_T\ll p_\text{max}$ we can neglect it in our analysis. Therefore we can obtain $\gamma_\mathfrak{m}(x_F)$ from
\begin{equation}
  \gamma_\mathfrak{m}\approx  \gamma_\text{cm} \frac{p_\text{max}}{m_\mathfrak{m}} \qty( \sqrt{x_F^2 + \frac{m_\mathfrak{m}^2}{p_\text{max}^2} } + \beta_\text{cm} x_F )\,.
\end{equation}
This equation can be inverted to yield two branches $x^{(\pm)}_F(\gamma_\mathfrak{m})$ 
\begin{equation}\label{feynman-x}
  x_F^{(\pm)}= -\gamma _{\text{cm}} \gamma _{\mathfrak{m}} \left(\beta _{\text{cm}}\pm \beta _{\mathfrak{m}}\right) \frac{m_{\mathfrak{m}} }{p_{\text{max} }}
\end{equation}
corresponding to the two solutions of the quadratic equation.
%$x_F$ can be negative (longitudinal momentum opposite to the COM boost vector, and if you do not include both branches you screw up the normalization.

\subsection{Meson decay to millicharged particles}
\label{apps:mesdec}

Most of the decay modes we consider involve two-body final states. For example, in the case of the 
$J/\psi$ the differential decay in the rest frame of the parent meson is mono-energetic 
$\dd \Gamma / \dd \mathcal{E}_\chi \propto \delta(\mathcal{E}_\chi - \tfrac12 m_{J/\psi})$ 
(in this subsection, curly letters refer to meson rest-frame quantities). Upon boosting to the lab frame 
this becomes a box distribution, $\text{Box}(E_\chi|\gamma_\mathfrak{m})$, of width 
$ E_\chi^{(+)}-E_\chi^{(-)}$ and height $1/(E_\chi^{(+)}-E_\chi^{(-)})$, where 
\begin{equation}
  E_\chi^{(\pm)}= \gamma_{J/\psi}(\mathcal{E}_\chi \pm \beta_{J/\psi} \mathcal{P}_\chi).
\end{equation}
Equivalently, in terms of the MCP's lab frame boosts, we have
\begin{equation}
  \gamma_\chi^{(\pm)}= \gamma_{J/\psi} \tilde{\gamma}_\chi (1 \pm \beta_{J/\psi} \tilde{\beta}_\chi)\,,
\end{equation}
where $\tilde{\gamma}_\chi$ and $\tilde{\beta}_\chi$ are the boost and velocity of the MCP in the meson 
rest frame.
In the case of $\rho^0$ and $\phi$, the dominant decay mode is also a two body final state  
(e.g.\ $\rho^0 \rightarrow \chi\bar{\chi}$). For $\omega$ the SM branching ratio for 
$\omega\rightarrow \pi^0 \ell^+ \ell^-$ is roughly ten times larger than 
$\omega\rightarrow \ell^+\ell^-$~\cite{Tanabashi:2018oca}, but this decay mode is only accessible for 
$m_\chi \leq \frac12 (m_\omega-m_\pi)\approx 325$ MeV, as opposed to 
$m_\chi \leq \tfrac12 m_\omega\approx 390$ MeV for the direct two body decay. We therefore 
neglect this decay mode\footnote{%
Including $\omega \rightarrow \pi^0 \chi \bar{\chi}$ would involve a chiral perturbation theory calculation analogous to the one performed for $\eta\rightarrow \gamma \chi\bar{\chi}$.
} 
which will underestimate the MCP flux by a factor of $\sim$ O(few) in the window 
$ 275\,\text{MeV} \gtrsim m_\chi \gtrsim 325 ~\text{MeV}$.  and focus instead on 
$\omega\rightarrow \chi\bar{\chi}$.   The branching ratio for MCPs can be obtained by a simple 
re-scaling of the di-muon branching ratio, 
\begin{equation}
    \text{BR}(\mathfrak{m}\rightarrow \chi\bar{\chi}) = \epsilon^2 \sqrt{\frac{m_{\mathfrak{m}}^2-4 m_\chi^2}{m_{\mathfrak{m}}^2-4m_\mu^2}}\text{BR}(\mathfrak{m}\rightarrow \mu^+ \mu^-)\,.
\end{equation}
where $\text{BR}(\rho^0\rightarrow \mu^+ \mu^-)=4.55 \times 10^{-5}$, $\text{BR}(\omega\rightarrow \mu^+ \mu^-)=7.4 \times 10^{-5}$, and $\text{BR}(\phi\rightarrow \mu^+ \mu^-)=2.87 \times 10^{-4}$ \cite{Tanabashi:2018oca}.

For the Dalitz decay $\eta\rightarrow \gamma \chi \bar{\chi}$ the MCPs are \emph{not} 
mono-energetic in the  meson rest frame. Nevertheless, each infinitesimal rest frame 
energy $\mathcal{E}_\chi$ can be treated as described above provided we integrate over 
all  such $\mathcal{E}_\chi$, weighted by the differential decay rate. Therefore, the lab-frame 
distribution of MCPs from $\eta$ decay is given by
\begin{equation}\label{eta-dist}
  \left[\frac{1}{\Gamma_\eta} \dv{\Gamma_\mathfrak{\eta}}{E_\chi}\right]_{\text{lab}}
  = \int \dd \mathcal{E}_\chi \left[\frac{1}{\Gamma_\eta} \dv{\Gamma_\eta}{\mathcal{E}_\chi}\right]_{\text{rest}} \times \text{Box}(E_\chi|\gamma_\mathfrak{m}) \,, 
\end{equation}
where $\Gamma_\eta= \Gamma(\eta\rightarrow \chi\bar{\chi}\gamma)$. From \cref{eta-dist} 
$P(\gamma_\chi|\gamma_\mathfrak{m})$ is readily obtained using \cref{p-gamma-chi} and the 
chain rule. We neglect the $\eta$ form factor and compute 
$\tfrac1\Gamma [ \dd \Gamma/\dd \mathcal{E}]$ using the Wess-Zumino $\gamma\gamma P$ 
vertex, with $P$ as pseudoscalar meson \cite{Pisarski:1997bq,Wess:1971yu}.

\section{Atmospheric meson production rate \label{app:meson-prod}}

Our treatment of meson production in the upper atmosphere is data driven and centers mostly around the 
ratio of $\sigma(pp\rightarrow \mathfrak{m} X)/\sigma_\text{inel}(pp)$ which varies as a function of 
center of mass energy. Although we have tried to inform our fits using data across a wide range of 
center of mass energies (or equivalently $\gamma_\text{cm}$) there is a limited window of ``important'' 
center of mass boosts that is determined by the competition between a rising inclusive cross section 
and a sharply falling cosmic-ray flux as a function of $\gamma_\text{cm}$ (the typical 
$I_\text{CR}\sim E^{-2.7}$ scaling translates to roughly  
$\mathcal{I}_\text{CR}\sim \gamma_\text{cm}^{-4.5}$). This is illustrated in \cref{production-modes} 
where we see that the relevant ranges are $\gamma_\text{cm}$ between 1.5-5 for $\eta$ mesons, 
between 1.5-10 for $\rho$ (and $\omega$) and $\phi$, and between 3-25 for $J/\psi$. %For clarity we have included a plot of $I_\text{CR}(\gamma_\text{cm})$ prior to folding against the production cross sections in \cref{ICR}. 

The rest of this section is devoted to our parameterization of the available inclusive cross section data, 
which we separate into a discussion of $\sigma_\mathfrak{m}(\gamma_\text{cm})$ and $P(\gamma_\mathfrak{m}|\gamma_\text{cm})$.  It is important to note that although  $P(\gamma_\mathfrak{m}|\gamma_\text{cm})$ is poorly constrained by the data we were able to find, its impact on our sensitivity curves is marginal; this is because the total number of MCPs produced is independent of this quantity. In contrast, although it has a relatively comprehensive dataset, the production cross section $\sigma_\mathfrak{m}(\gamma_\text{cm})$ in the window of maximal production (as shown in \cref{production-modes}) can have a substantial impact on the MCP signal (bounds on $\epsilon$ scale as $\sqrt[4]{\text{signal}}$) because it alters the total number of MCPs produced. We therefore anticipate that the uncertainties in the production cross section are the dominant source of error in our analysis (at the level of $\sim O$(few). 

\subsection{$\eta$ mesons} 

\begin{figure}
  \includegraphics[width=0.9\linewidth]{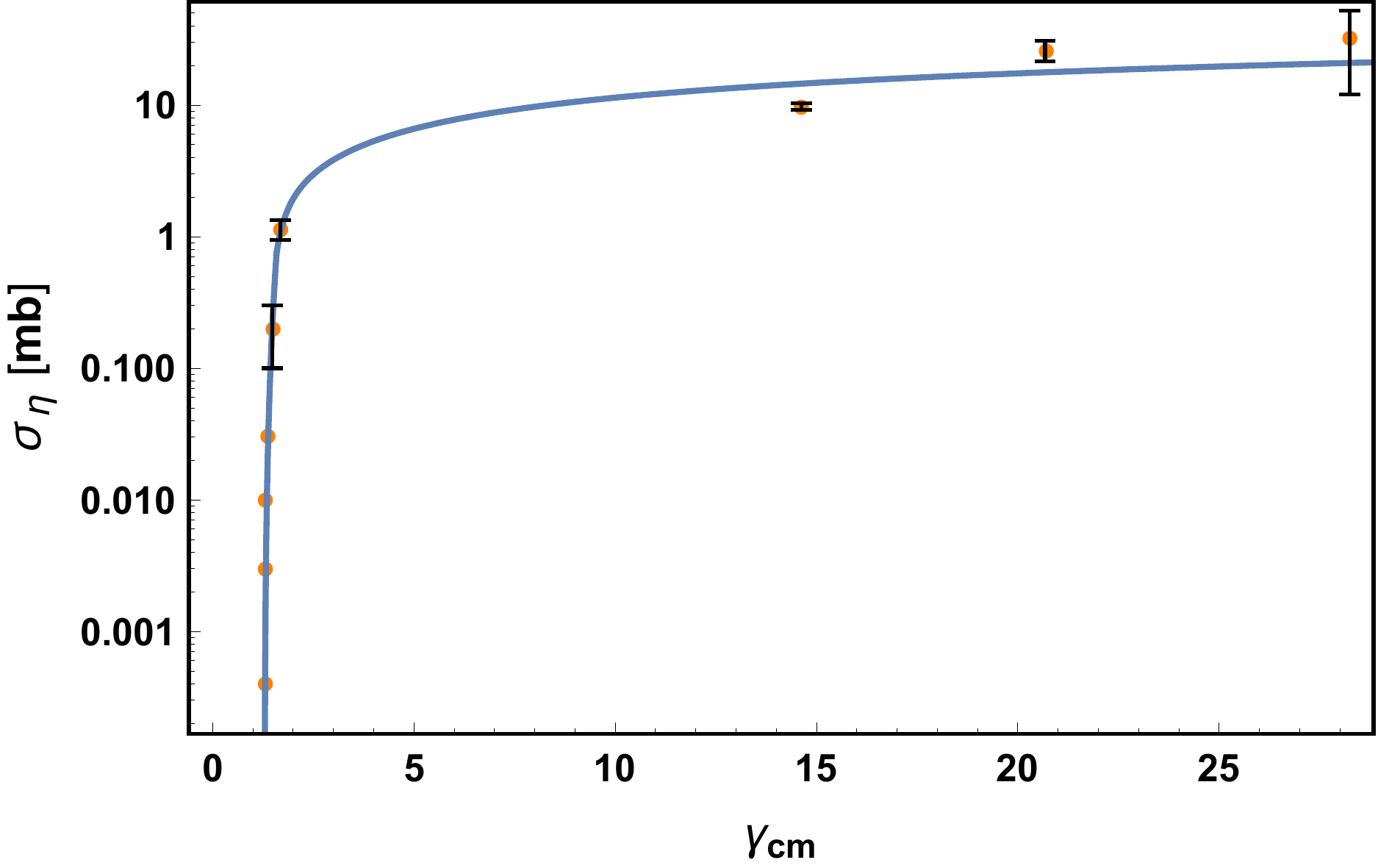}
  \caption{Production cross section for $pp\rightarrow \eta X$ as a function of 
  $\gamma_\text{cm}=\frac12\sqrt{s}/m_p$. The data is taken from 
  Refs.~\cite{HADES:2011ab,AguilarBenitez:1991yy,Jancso:1977dz,Hanlet:1995abs,Baldini:1988ti,Sibirtsev:1996ag} 
  and fitted using the piece-wise procedure described in the text; the smooth curve is 
  \cref{sigma-eta-fit}.   \label{Eta-meson}.}
\end{figure}

Eta meson production in $pp$ collisions has been most extensively measured in the near-threshold 
regime for the exclusive process $pp\rightarrow \eta pp$ \cite{HADES:2011ab,Sibirtsev:1996ag}. 
Near threshold this is the only available channel, such that this cross section can be taken as a 
reasonable estimate of the  total inclusive cross section. Further away from threshold bona fide 
measurements of the inclusive cross section are scarcer but we have identified four 
measurements in the literature at $\sqrt{s}=$3.17, 27.45, 38.8, and 
53\,GeV~\cite{HADES:2011ab,AguilarBenitez:1991yy,Jancso:1977dz,Hanlet:1995abs,Baldini:1988ti}. 
We split the available data into two subsets, near-threshold exclusive production (defined as 
$pp\rightarrow pp \eta$ measurements for $\sqrt{s}\leq 3$ GeV) and far-from-threshold inclusive 
data (defined as $pp\rightarrow \eta X$ for $\sqrt{s}> 3$ GeV). We fit the near-threshold data for 
$\sigma_\eta(\sqrt{s})$ with the function $f(x)=a (x - 2.42)^b x^{c}$ where 
$x\equiv\sqrt{s_{pp}/\text{GeV}}$. For the far-from-threshold data we instead use 
$g(x)=a(1+|b|/(x-2.42)^2)\log^2(x)$. In both cases a weighted linear regression to the data 
was performed.  Using the best fit values for both fits, and demanding that the function is 
continuous  we find
%have the following expression for $\sigma_\eta(\gamma_\text{cm})$
%
\begin{equation}\begin{split}
  \sigma_\eta(\gamma_\text{cm})= &\Theta(\gamma_\text{cm}-\gamma')f\qty(1.876 \gamma_\text{cm}) \\
  &+ \Theta(\gamma'-\gamma_\text{cm})  g\qty( 1.876\gamma_\text{cm})\,,
\end{split}\label{sigma-eta-fit}
\end{equation}
where the numerical factor comes from the relationship 
$\sqrt{s}=2m_p \gamma_\text{cm}= (1.876\,\text{GeV})\, \gamma_\text{cm}$. The functions 
$f(x)$ and $g(x)$, with their best fit values, are given by 
\begin{align}
  f(x)&=\qty(0.0176\,\text{mb}) \times (x-2.42)^{2.22} x^{4.59}\\
  g(x)&= \qty(1.32\,\text{mb}) \log ^2(x) \times \qty(1+\frac{0.356}{(x-2.42)^2})^{-1}\,,
\end{align}
and $\gamma'=1.59$ is chosen such that \cref{sigma-eta-fit} is continuous; the fit is shown vs.~the 
data (with error bars when available) in \cref{Eta-meson}. 

For the differential cross section $\dd \sigma_\mathfrak{m}/ \dd x_F$, measurements at 
NA27~\cite{AguilarBenitez:1991yy} strongly suggest an exponential distribution, 
\begin{equation}
  \dv{\sigma_\eta}{ x_F} = \sigma_\eta \times \frac{ c_\eta/2}{1 - \exp[-c_\eta] } \exp\qty[- c_\eta |x_F| ],\label{eta-exp}
\end{equation}
where $c_\eta$ depends on $\gamma_\text{cm}$. Measurements from NA27 at $\sqrt{s}=27.5$ GeV (corresponding to $\gamma_\text{cm}=14.6$)
fix $c_\eta\approx 9.5$ \cite{AguilarBenitez:1991yy}. One generally expects that $c_\eta$ will be a monotonically increasing function of  $\gamma_\text{cm}$, and that $c_\eta>0$. The simplest functional form that satisfies these expectations, and agrees with the measurement of \cite{AguilarBenitez:1991yy} is 
\begin{equation}\label{c-eta-guess}
  c_\eta(\gamma_\text{cm}) = 9.5 + (\text{slope})\times (\gamma_\text{cm}-14.6)\,;
\end{equation}
we take slope$\approx \tfrac12$. We checked that our sensitivity to MCPs from experiments such as SK are relatively insensitive to the value of the slope parameter.  

%%%%%%%%%%%%%%%%%%%%%%%%%%%%%%
%%%%%%%%%%%%%%%%%%%%%%%%%%%%%%
%%%%%%%%%%%%%%%%%%%%%%%%%%%%%%

\subsection{Light vector mesons}

\begin{figure}[t]
    \includegraphics[width=\linewidth]{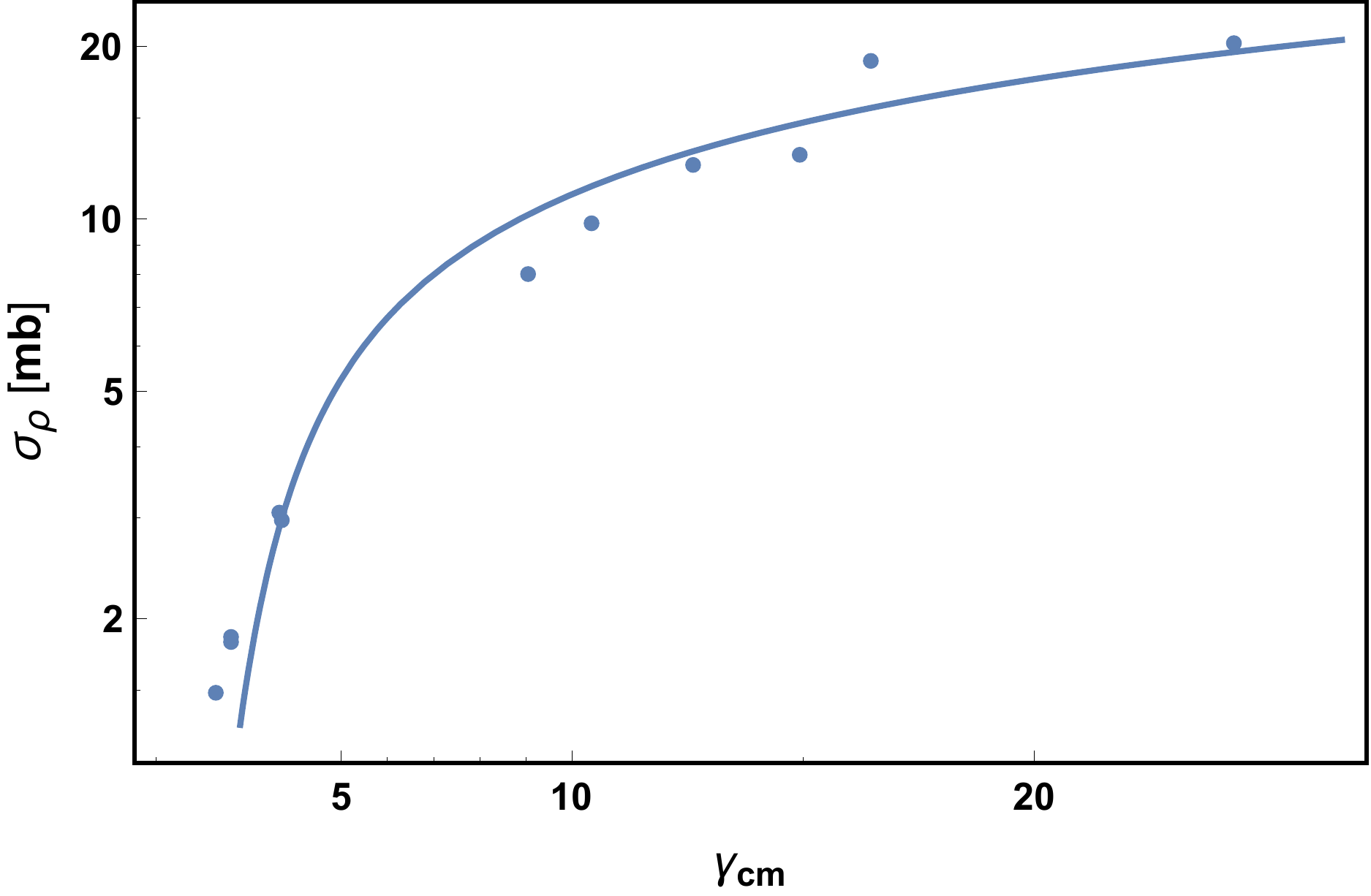}
    \caption{Compilation of $pp\rightarrow \rho X$ cross sections as a function of $\gamma_\text{cm}$ 
    taken from \cite{HADES:2011ab,Baldini:1988ti,Sibirtsev:1996ag} \label{rho-prod}. } 
\end{figure}

The production cross section for the $\rho^0$ meson is relatively well 
measured~\cite{HADES:2011ab,Baldini:1988ti,Sibirtsev:1996ag}. Like the $\eta$ meson we 
perform a best fit analysis with the function $g(x)$, but without weighted errors. We find the data to 
be reasonably well described by
\begin{equation}\begin{split}
    \sigma_\rho(\gamma_\text{cm})\approx (&1.35~\text{mb})\log^2(1.876 \gamma_\text{cm} )\\
    &\times \qty(1 + \frac{13.4}{(1.876 \gamma_\text{cm}-2.61)^2})^{-1}\,.
\end{split}\end{equation}
A comparison between the available data and our smooth fit is shown in \cref{rho-prod}. 

We found that the data for $\sigma(pp \rightarrow \rho X)$ had a much better coverage than the 
corresponding $\omega$ production cross section, and where there are measurements of the $\omega$ 
cross section it is nearly identical to the $\rho$ cross section. We therefore estimated the $\omega$ cross section  $\sigma_\omega(\gamma_\omega)\simeq\sigma_{\rho}(\gamma_\rho)$. 
%$\Phi_\omega(\gamma_\omega)=\Phi_{\rho}(\gamma_\rho)$. 

\begin{figure}[t]
    \includegraphics[width=\linewidth]{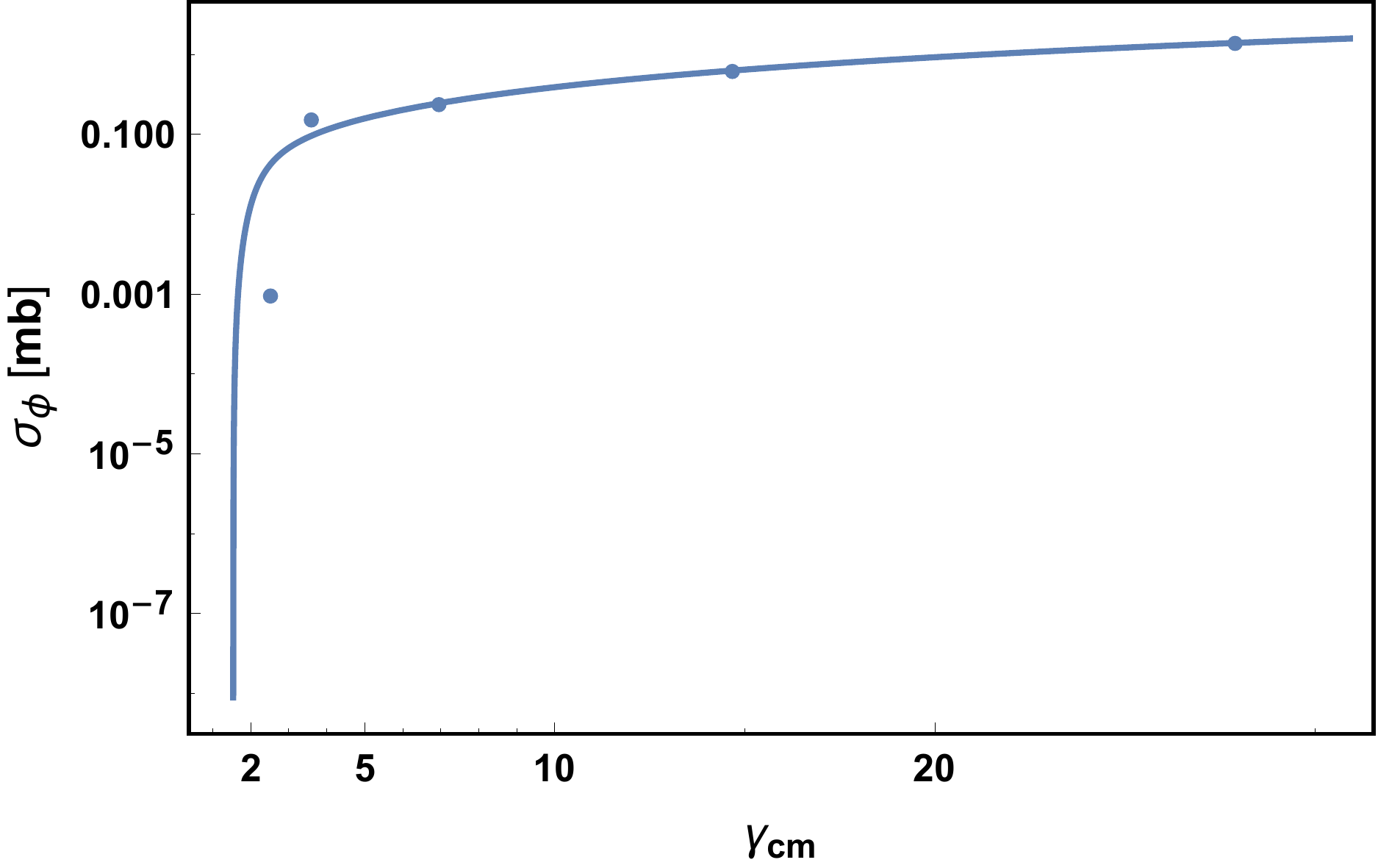}
        \caption{Compilation of $pp\rightarrow \phi X$ cross sections as a function of $\gamma_\text{cm}$ 
        taken from \cite{Baldini:1988ti,AguilarBenitez:1991yy, Sibirtsev:1996ag}. \label{phi-prod}} 
\end{figure}

For the $\phi$ meson we find that the functional form $h(x)=a(1+|b|/(x-2.896)^2)^{-1}x^c$  gives a 
reasonable fit to the data~\cite{HADES:2011ab,Baldini:1988ti,Sibirtsev:1996ag}. After an 
unweighted regression we find that $\sigma_\phi$ is well described, c.f.~\cref{phi-prod}, by 
\begin{equation}\begin{split}
    \sigma_\phi(\gamma_\text{cm})= (&0.01~\text{mb})(1.876 \gamma_\text{cm})^{1.23}\\
    & \times \qty(1+ \frac{2.4}{(1.876\gamma_\text{cm}-2.896)^2})\,. 
\end{split}\end{equation}

Like the $\eta$ meson, the longitudinal momentum distributions for the vector mesons were more 
difficult to find in the literature, and we rely on a single measurement at 
$\sqrt{s}=27.5$\,GeV~\cite{AguilarBenitez:1991yy} which shows the $x_F$ dependence to be 
described by \cref{eta-exp} with $c_V=c_\rho=c_\omega=c_\phi\approx 7.7$. We expect this 
value to be smaller at lower center of mass energies and so take
\begin{equation}\label{c-vector-guess}
    c_V= 7.7 + \frac{5.7}{13}(\gamma_\text{cm}-14.6)\,,
\end{equation}
which, just like the $c_\eta$, should be viewed as a cartoon of the behaviour of $\dd \sigma/\dd x_F$ as a function of $\gamma_\text{cm}$ rather than a faithful representation. 

\subsection{J/$\psi$ mesons}
\begin{figure}[t]
  \includegraphics[width=\linewidth]{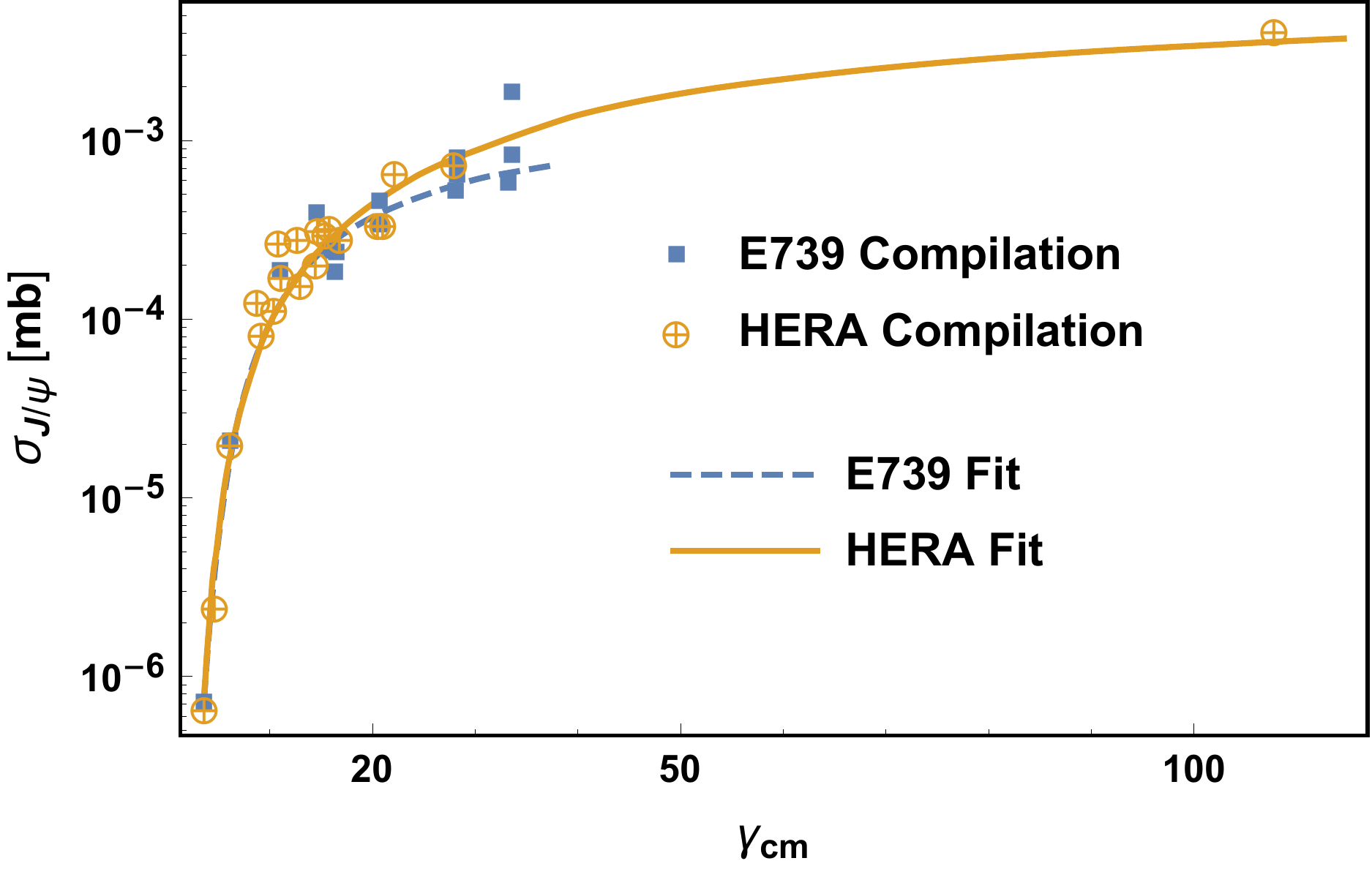}
  \caption{Production cross section for $pp\rightarrow J/\psi\, X$ as a function of 
  $\gamma_\text{cm}=\frac12\sqrt{s}/m_p$. The data points have been digitized from E-789's 
  compilation \cite{Kaplan:1996tb} 
  and HERA-B's compilation \cite{Abt:2005qr}. The solid curve is digitized from the fit presented in 
  Ref.~\cite{Abt:2005qr} while the dashed curve is taken from Ref.~\cite{Kaplan:1996tb}. For our 
  sensitivity analysis we use the solid curve from \cite{Abt:2005qr}. \label{j-psi-prod}}
\end{figure}

For the $J/\psi$ mesons we found two convenient summaries of the available data: one from 
E-739 (Fig.~7 in Ref.~\cite{Kaplan:1996tb}) and one from HERA-B (Fig.~8 of Ref.~\cite{Abt:2005qr}). 
The HERA-B compilation includes measurements at significantly higher center of mass energies. 
For comparison we plot both sets of data in \cref{j-psi-prod}  where we see that the HERA-B 
compiled data is roughly consistent with that from the E739 paper, but suggests a steeper 
growth with rising center of mass energy. We use the best fit to the  former to 
calculate $\sigma_\mathfrak{m}(\gamma_\text{cm})$. 

For the differential distribution we used the standard parameterization of 
$\dd \sigma_{J/\psi}/ \dd x_F$~\cite{Vogt:1999cu}
\begin{equation}\label{diff-jpsi}
  \dv{\sigma_{J/\psi}}{x_F} = \sigma_{J/\psi} \times \frac{(c_{J/\psi}+1)}{2}(1-|x_F|)^{c_{J/\psi}}\,.
\end{equation}
Like $c_\eta$, the fit parameter $c_{J/\psi}$ depends on the center of mass energy, and like $c_\eta$ 
the precise value of $c_{J/\psi}$ has a relatively mild effect on the fast-flux of MCPs. Data from 
experiments at lower energies  show a preference for $c_{J/\psi}\approx 2$ for 
$\sqrt{s}\leq 15$\,GeV~\cite{Vogt:1999cu} whereas experiments at higher energies find larger values 
such as  $c_{J/\Psi}\approx 6$ for $\sqrt{s}\approx 40$ GeV \cite{Alexopoulos:1997yd}; we did not 
find a robust set of measurements of $c_{J/\psi}$ spanning the entire range of $\gamma_\text{cm}$ 
relevant for cosmic ray $pp$ collisions.  For simplicity, and because our final results are relatively 
insensitive to the details of the $x_F$ distribution, we take $c_{J/\psi}$ to vary linearly with 
$\gamma_\text{cm}$,
\begin{equation}\label{c-j-psi}
  c_{J/\psi}=2 + \frac15(\gamma_\text{cm}-5)\,.
\end{equation}
In this case, there is data at lower center of mass energies that suggests this formula is a 
reasonable interpolation. 

\section{MCP signal in experiments \label{app:MCPsignals}
}

\subsection{MCP-electron scattering \label{app:mcpelecscat} }

\begin{figure}
  \includegraphics[width=0.9\linewidth]{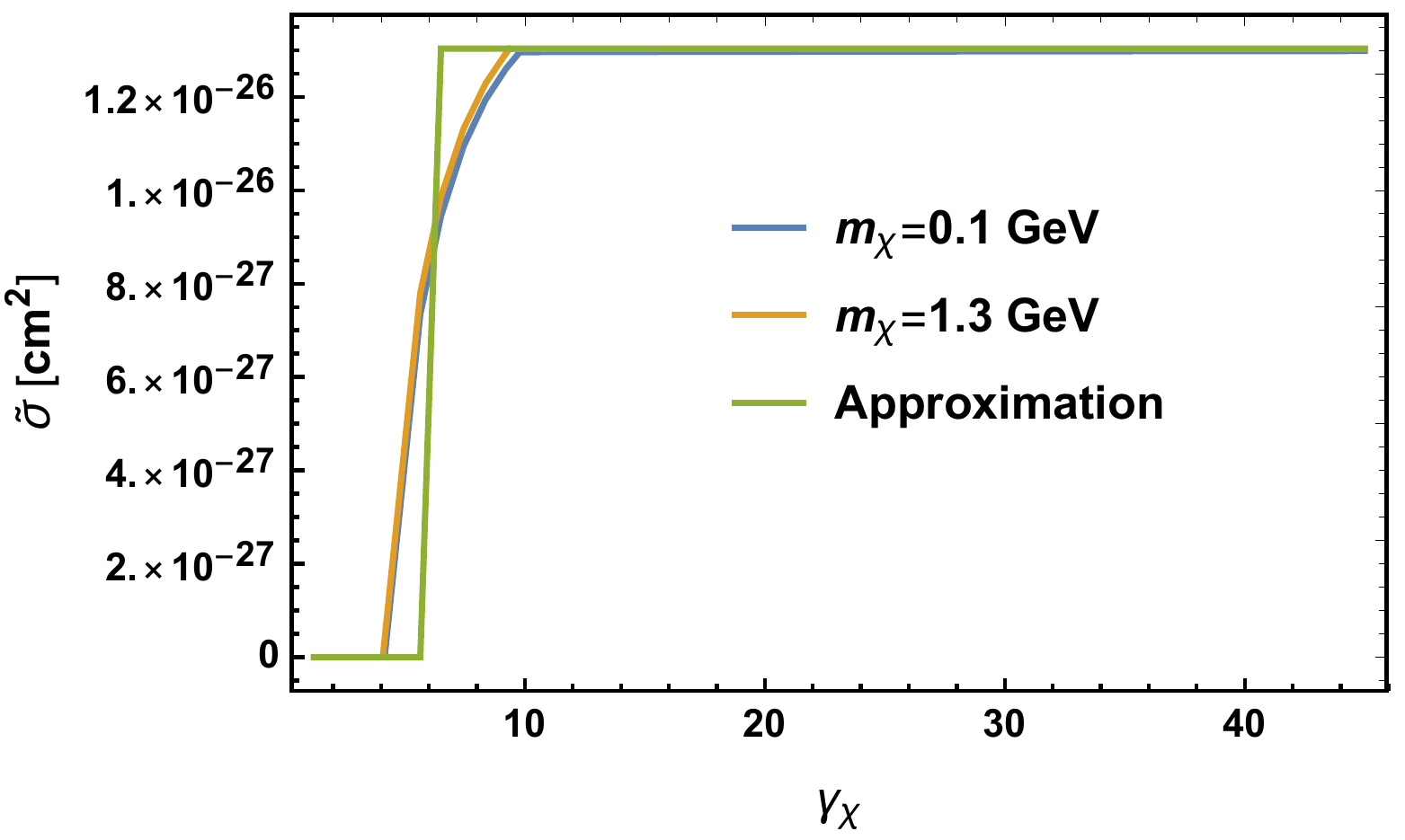}
  \caption{Dependence of windowed cross section $\tilde{\sigma}_{e\chi}$ on MCP boost factor $\gamma_\chi$ for $T'_\text{min}=16$ MeV and $T'_\text{max}=80$ MeV as compared to the approximation \cref{sigma-approx}. \label{cross-section-plot}  }
\end{figure}

The detection of MCPs is dominated by soft scattering from electrons as can be readily 
understood by considering the differential scattering cross section which, being mediated by photon 
exchange,  scales as $\dd\sigma / \dd Q^2 \sim 1/Q^4$. For elastic scattering from a target of mass 
$M$, the momentum transfer is given by $Q^2= 2 M(E'-M)$, where $E'$ is the total recoil energy of 
the target. The cross section is therefore maximized by scattering off the lightest target possible 
with the lowest possible recoil energy. In practice, experimental considerations such as detection 
efficiency and background reduction will set a minimum electron recoil energy which will, in turn, 
dictate the detection cross section for that given experiment. We will therefore consider a 
windowed cross section for electron recoils with kinetic energy, $T_e'=(E_e'-m_e)$ between 
$T_\text{min}$ and $T_\text{max}$, or equivalently with momentum transfers between 
$Q^2_\text{min}$ and $Q^2_\text{max}$, 
\begin{equation}\label{sigma-tilde}
  \tilde{\sigma}_{e\chi}= \int_{Q^2_\text{min}}^{Q^2_\text{max}} \frac{\dd \sigma_{e\chi} }{\dd Q^2}  \dd Q^2 \,.
\end{equation} 
Since the four-momentum transfer is directly related to the recoil energy in the lab frame, 
$T_e' = E_e' - m_e$,  via  $Q^2= -2 (p_e -p_e')^2= (2 m_e E_e' - 2m_e^2)= 2 m_e T_e'$, 
this is equivalent to demanding that $Q^2\geq 2 m_e T_\text{min}$. In the center of mass 
frame the maximal momentum transfer is given when the scattering is back-to-back such that 
$Q^2\leq  4 \mathcal{P}_e^2 -2 m_e^2$ where $\mathcal{P}_e$ is the electron's momentum 
in the center of mass frame, 
\begin{equation}
   \mathcal{P}_e =\sqrt{ \frac{m_e^4-2 m_e^2 \left(m_\chi^2+s\right)+\left(m_\chi^2-s\right)^2}{4s}} 
\end{equation}
In terms of lab frame variables this implies that 
\begin{equation}\label{Te-bound}
    T_e'  \leq \frac{2 m_e P_\chi^2}{2m_eE_\chi+m_e^2+m_\chi^2} \approx 2 m_e (\beta_\chi \gamma_\chi)^2
\end{equation}
where the approximation holds provided $m_\chi \gg m_e \gamma_\chi$. The main consequence 
of \cref{Te-bound} is that the lower bound of integration in \cref{sigma-tilde} is given (at leading order in 
$m_e/E_\chi$)\footnote{%
In producing our exclusion curves we use the full expression in \cref{Te-bound} rather than the 
indicated approximation.
}  
by $Q_\text{min}^2= \text{max}(2 m_e T_\text{min},  4 m^2_e (\beta_\chi \gamma_\chi)^2)$, and the 
upper bound is given by 
$Q_\text{max}^2 = \text{min}(2 m_e T_\text{max}, 4 m^2_e (\beta_\chi \gamma_\chi)^2)$.
The effect of this approximation on $\tilde{\sigma}_{e\chi}$
is stated as Eq.~(\ref{sigma-approx}) in the main text and illustrated in \cref{cross-section-plot}.
In summary, the primary driver of the windowed cross section is whether or not the incident MCP is 
sufficiently boosted to kick the electron above the detection threshold. In principle, the thresholds of 
large neutrino detectors can be rather low, a few MeV in case of SK, and as low as 200 keV for 
Borexino. We choose, however, a much higher threshold of $\sim 15-16$\,MeV, that removes all 
the events generated by solar neutrinos, so that background counting rates 
reduce to $O({\rm few})$ per year.

In the case of $e\chi$ scattering  the event shape spectrum is determined by the differential 
cross section with respect to recoil energy, $\dd \tilde{\sigma}/\dd T_e\propto 1/T_e^2$, and the 
incident flux of MCPs. We have confirmed that this shape is %qualitatively 
very similar to the case of a 
neutrino spectrum described by temperature of $T_\nu \lesssim 5$ MeV~\cite{Bays:2011si}, 
allowing us to readily employ those results.  

\begin{figure}[t]
    \includegraphics[width=\linewidth]{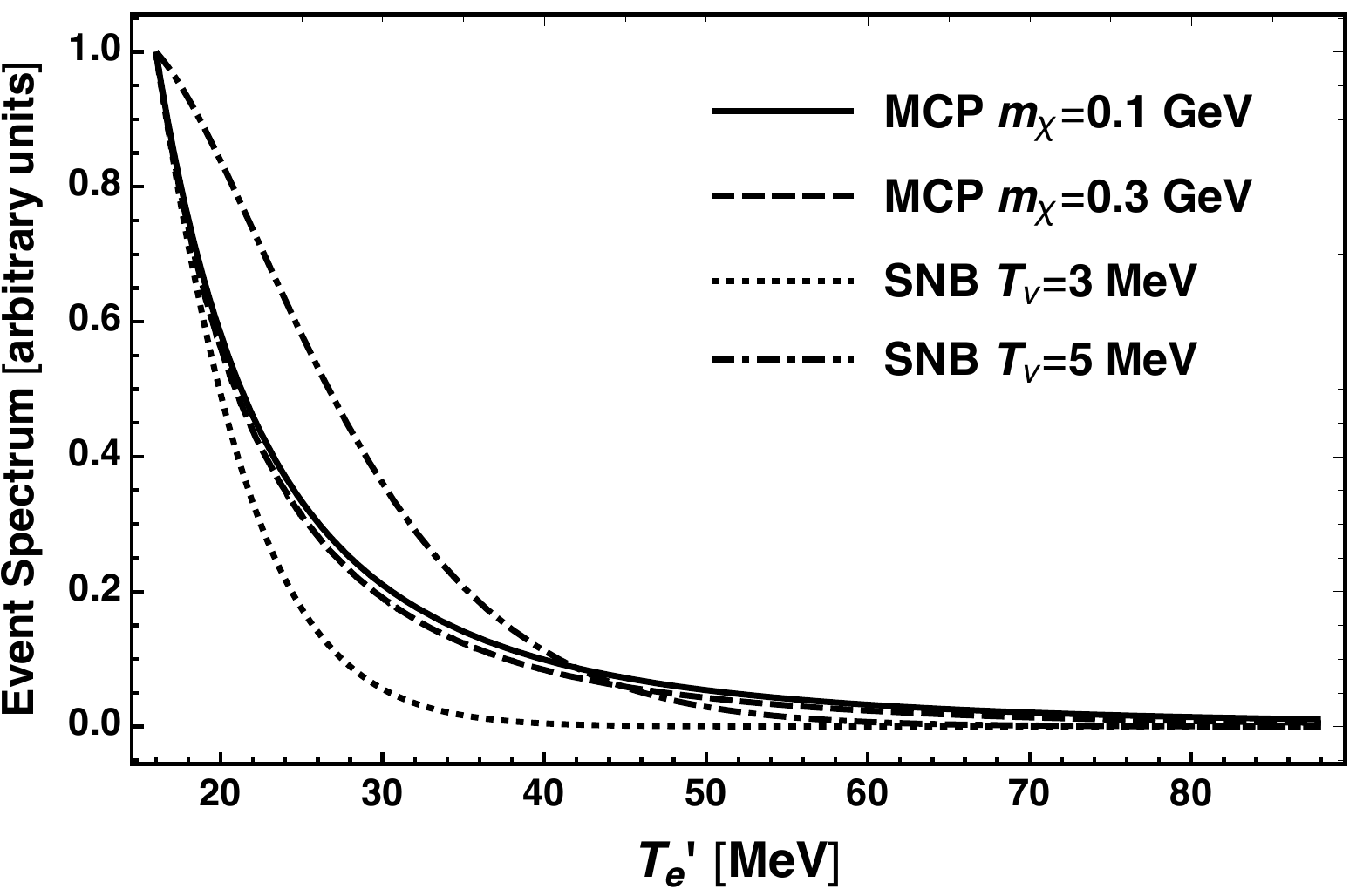}
    \caption{ Comparison of event shapes for MCP elastic scattering off electrons and inverse beta 
    decay from supernova background neutrinos. The MCP signal was obtained by folding the 
    differential scattering cross section $\dd \sigma_{e\chi}/ \dd T_e$ against the cosmic-ray 
    induced MCP flux. The supernova background curves correspond to $E_\nu^2/(\e^{E_\nu /T_\nu}+1)$ 
    where $E_\nu=T_e + 1.3$ MeV; these correspond to the fixed temperature profiles used 
    in Ref.~\cite{Bays:2011si} as can be readily verified by reproducing their Fig.\ 19. 
    \label{SNB-vs-MCP} }
\end{figure}

We emphasize that the DSNB limits from Super-K are given in terms of limits on the scattered 
positron event rate as a function of the effective neutrino temperature $T_{\nu}$ from supernova 
emission.  We note, for the reader's convenience, that in \cite{Bays:2011si} there are two bounds 
quoted: one for an ensemble of supernovae of different temperatures and one for a single 
supernova temperature. We use the latter, because it more closely mimics our signal as is 
clearly shown in \cref{SNB-vs-MCP} (the diffuse ensemble would be relatively flat as a function of energy).

\subsection{Ionization experiments}
 \label{apps:ionization}

Ionization is a very low threshold process and so we use the full flux (integrated over all boosts 
$\gamma_\chi$) of MCPs for ionization experiments; this corresponds to $\gamma_\text{cut}=1$ 
as shown in \cref{fastflux}; we denote this total flux by $\Phi(m_{\chi})$. To translate existing 
bounds on an ambient MCP flux in the literature we demand that 
\begin{equation}
    \epsilon^2 \times \qty( \epsilon^{-2} \Phi(m_\chi) ) = \Phi_\text{ion}(\epsilon)\,,
\end{equation}
where $\epsilon^{-2} \Phi_(m_\chi)$ corresponds to the $\gamma_\text{cut}=1$ curve 
in \cref{fastflux} (i.e.\ the integrated MCP flux generated in the upper atmosphere), and $\Phi_\text{ion}(\epsilon)$ is the joint exclusion curve obtained by combining 
data from MACRO \cite{Ambrosio:2000kh,Ambrosio:2004ub} and Majorana \cite{Alvis:2018yte} as 
shown in Fig.\ 7 of \cite{Alvis:2018yte}. We then solve for $\epsilon$ for each value of 
$m_\chi$ which determines a critical value of,  $\epsilon_c(m_\chi)$, above which MCPs are excluded. %Our analysis  does not model attenuation due to the overburden above either experiment. For MACRO, the bounds on the flux are extremely strong (roughly six orders of magnitude smaller than the calculated fast-flux at $\epsilon=0.2$). The reason that the bound in \cref{money-plot is set at $\epsilon\lesssim 0.2$ is that the curve published by MACRO terminates at $\epsilon \sim 0.2$. Therefore, the bounds obtained from MACRO are robust even if only about one part per million of the cosmic-ray generated MCP spectrum penetrates the overburden. For Majorana, as argued in \cref{sec:CR_meson}, attenuation is likely not an issue provided $\epsilon_c(m_\chi) \lesssim 0.1$.  }

\subsection{Meson fluxes}

\begin{figure}
    \includegraphics[width=\linewidth]{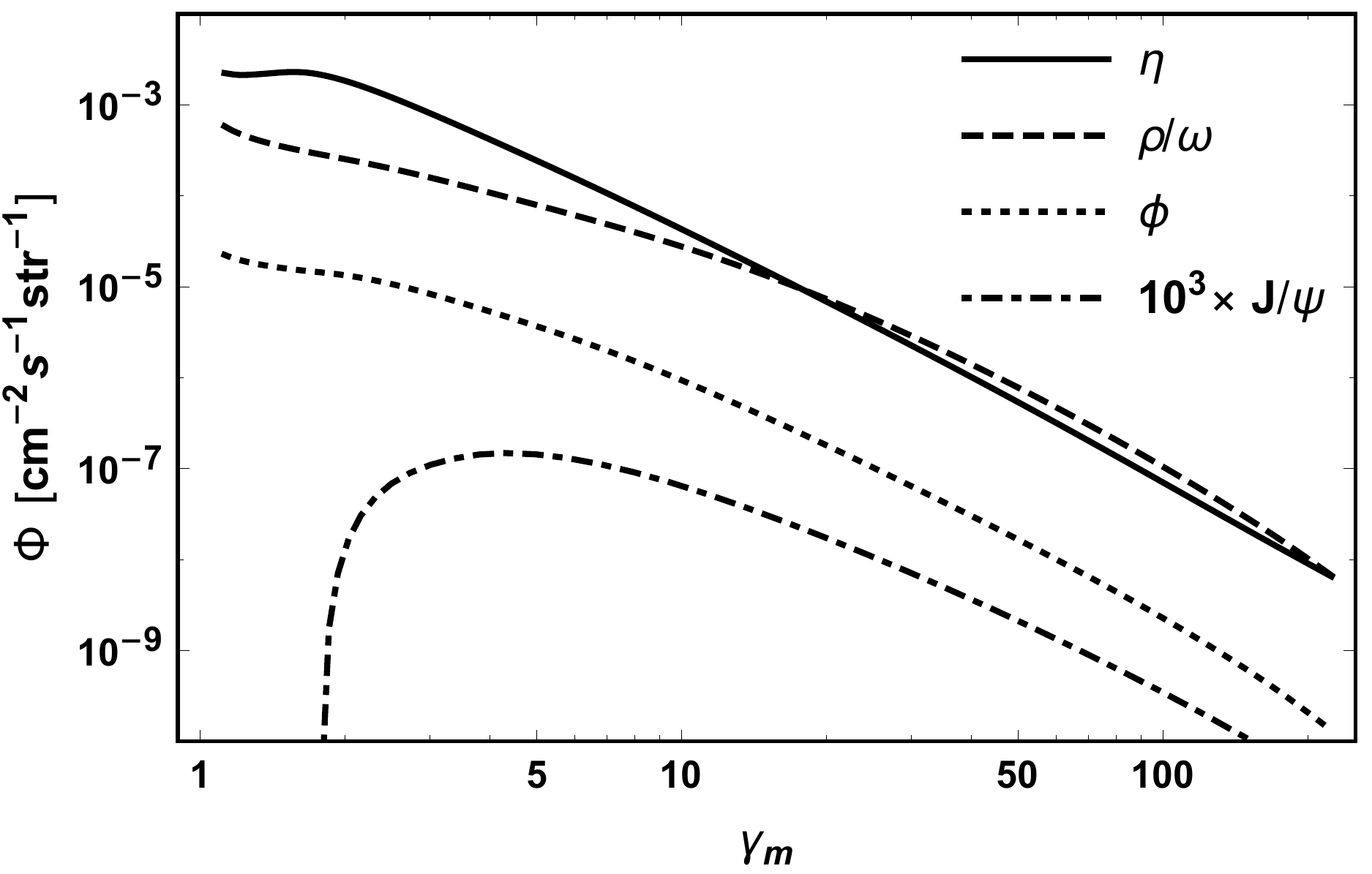}
    \caption{Mesons fluxes from primary $pp$ collisions assuming a longitudinal momentum 
    distribution as described in \cref{c-j-psi,diff-jpsi,c-vector-guess,eta-exp,c-eta-guess}. \label{meson-fluxes} }
\end{figure}

A useful biproduct of our research are the lab-frame spectra of mesons as a function of 
$\gamma_\mathfrak{m}$. Given any calculable $\mathfrak{m}\rightarrow \text{dark sector}$ decay, 
using the meson spectra as inputs, a flux of dark sector particles originating from primary cosmic-ray collisions can be obtained. Our results, shown in \cref{meson-fluxes}, rely only on simple 
parameterizations of the differential cross sections $\dd \sigma / \dd x_F$ and the measured production cross sections as outlined 
in Appendix \ref{app:meson-prod}. 
\vfill 

\pagebreak 

$~~~~$ 

\vfill

\bibliography{MCP_Refs.bib}
\end{document}